\documentclass[twocolumn]{aastex63} 
\usepackage[flushleft]{threeparttable}
\usepackage{hyperref}
\usepackage{longtable}
\usepackage{tabularx}
\usepackage{amsmath,amstext}
\usepackage[figure,figure*]{hypcap}
\usepackage{newtxmath} 
\usepackage{makecell}
\usepackage{gensymb}

\usepackage{lineno}

\shorttitle{Metal-Poor Stars Observed with SALT}
\shortauthors{Zepeda et al.}

\submitjournal{ApJ}
\accepted{November 15, 2021}

\DeclareUnicodeCharacter{2212}{-} 
\begin{document}

\title{Metal-Poor Stars Observed with the Southern African Large Telescope II. An Extended Sample
} 

\correspondingauthor{Joseph Zepeda}
\email{jzepeda1@nd.edu}
\author{Joseph Zepeda}
\affiliation{Department of Physics, University of Notre Dame, Notre Dame, IN 46556, USA} 
\affiliation{JINA Center for the Evolution of the Elements (JINA-CEE), USA}

\author{Kaitlin C. Rasmussen}
\affiliation{Department of Astronomy, University of Michigan, 1085 S. University Ave., Ann Arbor, MI 48109, USA} 
\affiliation{JINA Center for the Evolution of the Elements (JINA-CEE), USA}

\author{Timothy C. Beers} 
\affiliation{Department of Physics, University of Notre Dame, Notre Dame, IN 46556, USA} 
\affiliation{JINA Center for the Evolution of the Elements (JINA-CEE), USA}

\author{Vinicius M. Placco}
\affiliation{NSF’s NOIRLab, 950 N. Cherry Ave., Tucson, AZ 85719, USA}
\affiliation{Department of Physics, University of Notre Dame, Notre Dame, IN 46556, USA}  
\affiliation{JINA Center for the Evolution of the Elements (JINA-CEE), USA}

\author{Yang Huang}
\affiliation{South-Western Institute for Astronomy Research, Yunnan University, Kunming 650500, People’s Republic of China}

\author{\'Eric Depagne} 
\affiliation{South African Astronomical Observatory (SAAO), Observatory Road Observatory Cape Town, WC 7925, South Africa}  

\begin{abstract}
We present results from high-resolution (R $\sim 40,000$) spectroscopic observations of over 200 metal-poor stars, mostly selected from the RAVE survey, using the Southern African Large Telescope. We were able to derive stellar parameters for a total of 108 stars; an additional sample of  50 stars from this same effort was previously reported on by Rasmussen et al..  Among our newly reported observations, we identify 84 very metal-poor (VMP; [Fe/H] $< -2.0$, 53 newly identified) stars and 3 extremely metal-poor (EMP; [Fe/H] $< -3.0$,  1 newly identified) stars.  The elemental abundances were measured for carbon, as well as several other $\alpha$-elements (Mg, Ca, Sc, Ti), iron-peak elements (Mn, Co, Ni, Zn), and neutron-capture elements (Sr, Ba, Eu). Based on these measurements, the stars are classified by their carbon and neutron-capture abundances into carbon-enhanced metal-poor (CEMP; [C/Fe] $> +0.70$), CEMP sub-classes, and by the level of their $r$-process abundances.  A total of 17 are classified as CEMP stars. There are 11 CEMP-$r$ stars (8 newly identified), 1 CEMP-$s$ star (newly identified), 2 possible CEMP-$i$ stars (1 newly identified), and 3 CEMP-no stars (all newly identified) in this work.  We found 11 stars (8 newly identified) that are strongly enhanced in $r$-process elements ($r$-II; [Eu/Fe] $> +0.70$), 38 stars (31 newly identified) that are moderately enhanced in $r$-process elements ($r$-I; $+0.30 < $ [Eu/Fe] $\leq +0.70$), and 1 newly identified limited-$r$ star.  

\end{abstract}


\section{Introduction}

The first stars, thought to have formed several hundred million years after the Big Bang, were born from gas comprising only hydrogen, helium, and trace amounts of lithium \citep{PhysRevLett.16.410, Wagoner_1967}.  As pointed out by numerous authors, pristine gas in the environment of first-star formation cools primarily via H$_2$ line cooling, which is significantly less efficient than cooling by heavier elements and dust.  In the absence of these cooling channels, it is expected that the first stars had a mass function that favored higher mass stars than the current IMF (\citealt{Abel93}, \citealt{Bromm_2002}; \citealt{Susa_2019}, and references therein). Unless the First Mass Function included stars with masses as low as $\sim 1 M_{\odot}$, then the first stars, often referred to as Population III (Pop III), had main-sequence lifetimes that were too short for direct observation and study in the Milky Way today. 

The explosive processes following the deaths of massive stars can eject any elements formed during their short evolution into the interstellar medium, which can mix with pristine gas (and the ejecta from other dying stars) and form new stars.  Cooling pathways that are opened permit the formation of second-generation (Population II; Pop II) stars with sufficiently low mass (and long lifetimes) that they should be presently observable \citep{Silk_1977, Sabano_1977, Bromm_2003, Frebel_2007, Frebel_2012, Susa_2014, Chiaki_2017, Dutta_2020}. Since the photospheres of low-mass stars preserve the composition of their natal gas (unless polluted by mass-transfer events from binary companions or other sources), the identification and observation of particularly low-metallicity Pop II stars is crucial for understanding the nature of the Pop III stars.

\citet{Beers_1992} first identified that a surprising fraction of the lowest metallicity stars discovered in the HK Survey exhibited strong CH $G$-band features.  Subsequent observations of the most metal-poor stars have demonstrated that, as metallicity decreases, the fraction of stars with large enhancements of carbon relative to iron indeed increases dramatically \citep{Rossi_1999, Rossi_2005, Lucatello_2006, Lee_2013, Placco_2014, Yoon_2018}, suggesting that carbon enhancement may be a primary signature of the first Pop II stars.  

\citet{Beers_2005} defined the chemically peculiar class of Carbon-Enhanced Metal-Poor (CEMP) stars, originally based on a carbon-abundance ratio [C/Fe] $ > +1.0$ (more recently lowered to [C/Fe] $> +0.7$ by \citealt{aoki2007}). Nucleosynthesis models for Pop III stars have shown the production of carbon-rich and iron-poor ejecta \citep{Meynet_2006, Nomoto_2006, Meynet_2010, Ishigaki_2014, Tominaga_2014, Choplin_2016, Kobayashi_2020} are in fact expected. Additionally, observations of damped Ly$\alpha$ systems have exhibited elemental abundances including carbon enhancement matching the predicted Pop III yields \citep{Cooke:2011au, 10.1111/j.1365-2966.2011.19365.x,Zou_2020}, but see also the arguments in \citet{Frebel_2015}.  Although further work is required, the current evidence points towards at least some CEMP stars being likely among the first Pop II stars to form.

Detailed studies of the elemental abundances for CEMP stars over the past few decades have revealed 
sub-populations that also provide information on the origin of the neutron-capture (n-capture) elements.  
There are three dominant processes identified for the production of n-capture elements, the slow ($s$), intermediate ($i$), and rapid ($r$) process, associated in turn with increasing levels of neutron flux density in their production sites (see \citealt{Cowan_2021} for a recent review). Numerous CEMP stars exhibit over-abundances of n-capture elements produced by these processes, originally sub-classified by \citet{Beers_2005} as CEMP-$s$, CEMP-$r/s$ (now often referred to as CEMP-$i$, see, e.g., \citealt{Hampel_2016, Purandardas_2021}), and CEMP-$r$ stars.  The CEMP-no stars (which do not exhibit  over-abundances of n-capture elements) have been shown to dominate the fractions of carbon-enhanced stars at the lowest metallicities (\citealt{Placco_2014, Yoon_2016}, and references therein), and may be the direct descendants of Pop III stars (see, e.g., \citealt{Chiaki_2020, Almusleh_2021}, and references therein).  High-resolution spectroscopic analyses for larger samples of all of these sub-classes would improve our understanding of their astrophysical origins.  Table~\ref{tab:classes} summarizes the current definitions of these sub-classes, based in part on recent revisions from \citet{Frebel_2018}. 

The origins of the various sub-classes of $r$-process-enhanced (RPE) stars, also listed in Table~\ref{tab:classes}, are the subject of active investigation. Neutron star mergers (NSMs) have gained support as sources of heavy $r$-process material, with the inference of lanthanide material synthesized by the kilonova associated with the gravitational wave signal from a NSM detected by LIGO,
GW170817 \citep{Abbott_2017, Drout_2017, Kilpatrick_2017, Shappee_2017}. In addition,
recent observations of RPE stars in the satellite ultra-faint dwarf (UFD) galaxies Reticulum II  \citep{Ji_2016, Ji_2019, Roederer_2016} and Tucana III \citep{Hansen_2017, Marshall_2019}, as well as in the Magellanic Clouds \citep{Reggiani_2021}, demonstrate that UFDs and dwarf spheroidal galaxies are the likely birthplaces of the metal-poor RPE stars in the halo of the Milky Way, which were then strewn throughout it when their parent systems were disrupted during accretion (see, e.g., \citealt{Roederer_2018}, \citealt{Brauer_2019}, and \citealt{Gudin_2021}).  Although rapid progress has been made with the dedicated $R$-Process Alliance survey (\citealt{Holmbeck_2020}, and references therein), further expansion of the numbers of RPE stars with high-resolution spectroscopic analyses is required in order to better constrain the possible variety of astrophysical origin of this process.  

This paper, the second which describes high-resolution spectroscopic observations of metal-poor stars with the Southern African Large Telescope (SALT), adds a number of newly analysed high-resolution spectra for CEMP and RPE stars, as well as for ``normal" metal-poor stars in the halo of the Milky Way. Section \ref{sec:style} describes the target selection and observations of our SALT sample. Section \ref{sec:reduction} explains the reduction and stellar parameter derivation processes. In Section \ref{sec:abun}, we describe the derivation of abundances for the stars in our sample. Section \ref{sec:result} reports the final elemental abundances and classifications of our stars.  In Section \ref{sec:sum}, we  summarize these results and future plans for these data.

\begin{deluxetable}{lr}
\tablecaption{\label{tab:classes} Metal-Poor Sub-class Definitions}
\tabletypesize{\scriptsize}
\tablehead{
\colhead{Sub-classes}  & 
\colhead{Definition}}
\startdata
CEMP & [C/Fe] $>$ +0.7\\
CEMP-$r$ &  [C/Fe] $>$ +0.7, [Eu/Fe] $>$ +0.70, [Ba/Eu] $<$ 0.0\\
CEMP-$s$ &  [C/Fe] $>$ +0.7, [Ba/Fe] $>$ +1.0, [Ba/Eu] $>$ +0.5\\
CEMP-$i$ ($r/s$) & [C/Fe] $>$ +0.7, 0.0 $<$ [Ba/Eu] $<$ +0.5 or 0.0 $\leq$ [La/Eu] $\leq$ +0.7\\
CEMP-no &  [C/Fe] $>$ +0.7, [Ba/Fe] $<$ 0 \\
$r$-II & [Eu/Fe] $> +0.70$, [Ba/Eu] $<$ 0.0 \\
$r$-I & $+0.30 < $ [Eu/Fe] $\leq +0.70$, [Ba/Eu] $<$ 0.0 \\
Limited-$r$ & [Sr/Ba] $> + 0.5$, [Eu/Fe] $\leq +0.3$ 
\enddata
\end{deluxetable}

\section{Target Selection and Observations} \label{sec:style}

The first paper in this effort \citep{Rasmussen_2020} outlined the target selection in detail; here we provide a brief summary.

\subsection{Target Selection}

\begin{figure}[t]
    \centering
    \includegraphics[scale = 0.32]{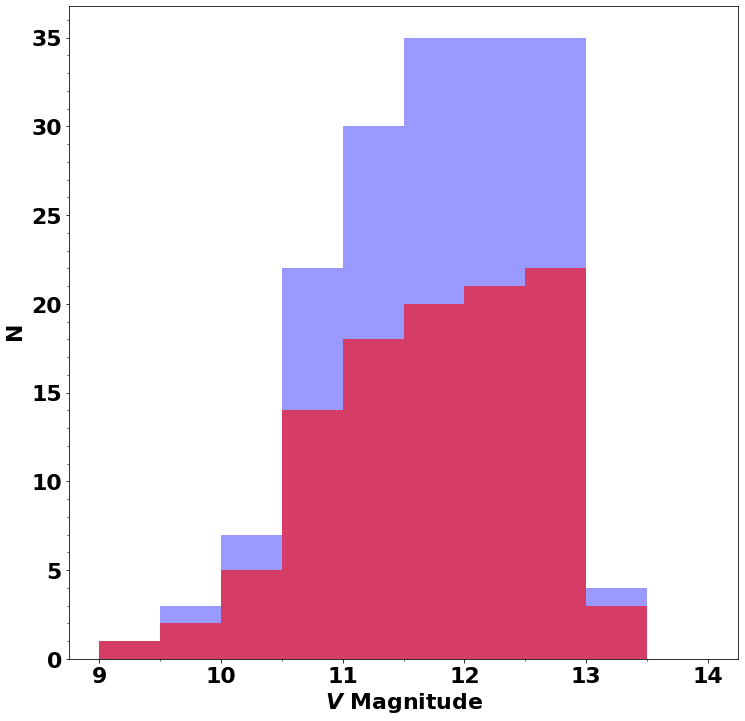}
    \caption{The \textit{V} magnitudes of the sample stars. Stars that were observed with SALT are shown in blue.  The subset of stars for which stellar parameters and abundance analyses could be carried out are shown in red.}
    \label{fig:maghist}
\end{figure}

The stars observed in this survey were selected for their brightness and low metallicity. Most of the candidates came from the Radial Velocity Experiment (RAVE), nominally with [Fe/H] $\lesssim -2.0$, however other stars with similarly low metallicities were also observed. This survey is magnitude limited ($9 < I < 12$), which alleviates some potential selection biases. The RAVE survey used measurements of the calcium triplet to estimate stellar parameters, including metallicities, for its stars (\citealt{Rave1}, \citealt{Rave5}).  It became clear, however, based on early follow-up observations of the metal-poor candidates from RAVE, that about 30-40\% of the these candidates in fact possessed metallicities that were somewhat higher than in the (early) published work, hence we initiated a more extensive follow-up campaign in order to validate RAVE candidates prior to obtaining additional high-resolution spectroscopic follow-up.
The metal-poor RAVE candidates were observed at medium resolution (1200 $\lesssim R \lesssim 2000$) with the following telescopes: the New Technology Telescope, Mayall Telescope, Gemini North, Gemini South, and SOAR. The spectra obtained with these telescopes where then processed with the non-Segue Stellar Parameter Pipeline (n-SSPP; see \citealt{Beers_2014,Beers_2017}) to obtain stellar parameters and [C/Fe] abundance ratios; for details see \citet{Placco_2018}. Note that the [C/Fe] obtained in \cite{Placco_2018} were corrected for the effects of carbon depletion during ascent of the giant branch, as described in \citet{Placco_2014}. All of these stars had \textit{V} magnitudes brighter than 14;  a histogram of the \textit{V} magnitudes for the target stars that were observed with SALT is shown in Figure \ref{fig:maghist}.  The figure shows the full sample of targets (including those previously reported in \citealt{Rasmussen_2020}), as well as the subset of stars for which stellar parameters and abundance analyses could be carried out for, as described below.

\subsection{SALT Observations}

A total of 223 high-resolution ($R \sim 40,000$) spectroscopic observations of these relatively bright targets were completed with SALT from April 2017 to April 2019, primarily during commissioning or bad-weather time.  The total observation time was 55 hours. However, the low signal-to-noise (S/N $<$ 20 at 5333\,{\AA}) for many of the targets prevented accurate stellar-parameter determination and abundance measurements\footnote{Note that \citet{Rasmussen_2020} measured S/N at 4300\,{\AA} while in this paper we measured S/N at 5333\,{\AA}, resulting in slightly higher reported S/N.}.   The High-Resolution Spectrograph (HRS) on SALT is a dual-beam, single-object, fiber-fed e\'chelle spectrograph, which yielded a usable wavelength range from 3970-5560\,{\AA}. The spectra had too low S/N below 3970\,{\AA}, and the pipeline used for reduction prevented use of the spectra redward of 5560\,{\AA}. The position of each target in degrees, the UT date of observation, and the radial velocity, along with its error, are listed in Table \ref{tab:rv}. 

\begin{deluxetable*}{ccccc}[t]
\tablecaption{\label{tab:rv} Position, Date of Observation, and Radial Velocity}
\tabletypesize{\scriptsize}
\tablehead{
\colhead{Name}  &
\colhead{RA} &
\colhead{Dec} &
\colhead{UT Date} &
\colhead{RV} 
\\
& \colhead{(degrees)} & \colhead{(degrees)} &  & \colhead{(km s$^{-1}$)}
}
\startdata
RAVE J000438.4-540134 & 	 +1.1583 & 	 $-$54.0261 & 	October 1 2018 & 	 +191.2$\pm$2.1 \\
RAVE J001236.5-181631 & 	 +3.1500 & 	 $-$18.2750 & 	July 4 2017 & 	 $-$121.8$\pm$1.3 \\
RAVE J003946.9-684957 & 	 +9.9458 & 	 $-$68.8331 & 	November 28 2018 & 	 +81.3$\pm$1.2 \\
RAVE J010839.6-285701 & 	 +17.1667 & 	 $-$28.9500 & 	July 7 2017 & 	 +124.6$\pm$3.6 \\
2MASS J01162885-5343408 & 	 +19.1208 & 	 $-$53.7281 & 	July 29 2018 & 	 $-$16.7$\pm$0.8 \\
2MASS J01201051-4849461 & 	 +20.0458 & 	 $-$48.8300 & 	October 21 2018 & 	 $-$64.1$\pm$7.9 \\
RAVE J012931.1-160046 & 	 +22.3792 & 	 $-$16.0131 & 	June 12 2017 & 	 +111.1$\pm$9.3 \\
RAVE J013807.0-195210 & 	 +24.5292 & 	 $-$19.8689 & 	October 21 2018 & 	 $-$66.1$\pm$3.1 \\
RAVE J014254.2-503249 & 	 +25.7250 & 	 $-$50.5469 & 	June 21 2017 & 	 +143.9$\pm$1.7 \\
RAVE J014908.0-491143 & 	 +27.2833 & 	 $-$49.1950 & 	June 13 2017 & 	 +72.6$\pm$2.8 \\
\enddata
\tablecomments{This table is a stub; the full table is available in the electronic edition.}
\end{deluxetable*}

\section{Data Reduction, Estimation of Stellar Parameters, and Comparisons with Previous Stellar-Parameter Estimates} \label{sec:reduction}

In this section, we briefly summarize the data reduction, derivation of the stellar parameters ($T_{\rm eff}$, $\log g$, and [Fe/H]), and a comparison to stellar parameters for stars in common with previous (lower-resolution) analyses.

\begin{figure}[t]
    \centering
    \includegraphics[scale = 0.32]{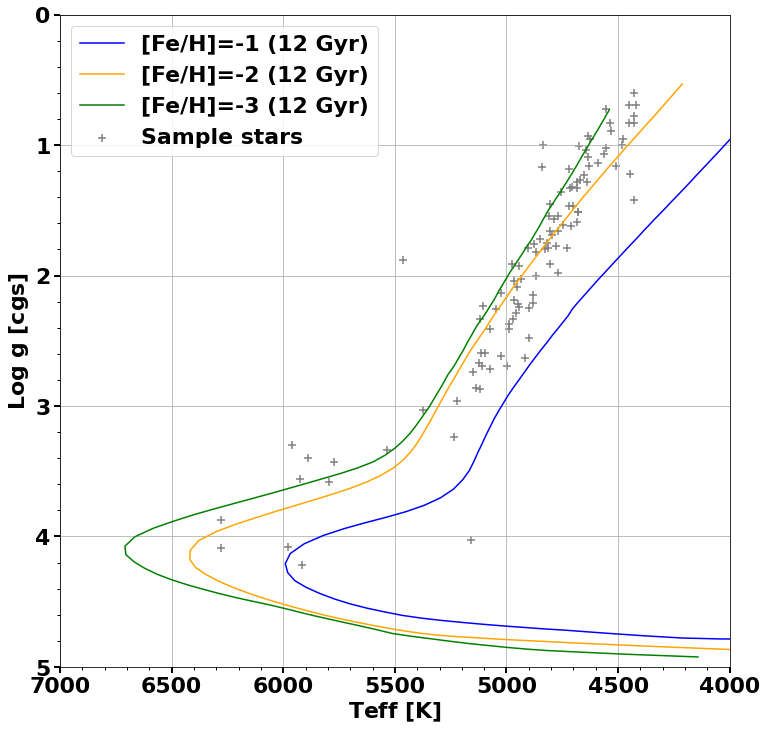}
    \caption{Isochrones for stars with derived stellar-parameter estimates.  The different colors represent isochrones at 12 Gyr with different metallicities, taken from \citet{Demarque_2004}}
    \label{fig:isochrone}
\end{figure}

\subsection{Data Reduction}

Data reduction was accomplished with an IRAF\footnote{IRAF was distributed by the National Optical Astronomy Observatory, which was managed by the Association of Universities for Research in Astronomy (AURA) under a cooperative agreement with the National Science Foundation.} pipeline created for these data, making use of the \texttt{noao, imred, cddred}, and \texttt{echelle} packages. The spectra (each from a single observation) were then normalized, stitched together, and radial-velocity corrected via a cross-correlation procedure using the software Spectroscopy Made Harder (SMHR; \citealt{Casey2014}). 

\begin{deluxetable*}{ccccccc}[t]
\tablecaption{\label{tab:sp}Signal-to-Noise and Stellar Parameters}
\tabletypesize{\scriptsize}
\tablehead{
\colhead{Name}  &
\colhead{Other Name} &
\colhead{S/N } & 
\colhead{$T_{\rm eff}$} &
\colhead{log $g$} &
\colhead{$\xi$} &
\colhead{[Fe/H]}\\
& & \colhead{(5033\,\AA)}&\colhead{(K)} & \colhead{(cgs)} & \colhead{(km s$^{-1}$)} & \colhead{} 
}
\startdata
RAVE J000438.4-540134 & 	--- & 	30 & 	4678$\pm$150 & 	1.51$\pm$0.30 & 	1.87$\pm$0.20 & 	$-$2.38$\pm$0.10 \\
RAVE J001236.5-181631 & 	--- & 	57 & 	5073$\pm$150 & 	2.41$\pm$0.30 & 	1.63$\pm$0.20 & 	$-$2.41$\pm$0.09 \\
RAVE J003946.9-684957 & 	--- & 	32 & 	4631$\pm$150 & 	1.16$\pm$0.30 & 	1.88$\pm$0.20 & 	$-$2.62$\pm$0.10 \\
RAVE J010839.6-285701 & 	SMSS J010839.58-285701.5 & 	27 & 	4843$\pm$150 & 	1.17$\pm$0.30 & 	1.61$\pm$0.20 & 	$-$2.87$\pm$0.14 \\
2MASS J01162885-5343408 & 	--- & 	32 & 	4808$\pm$150 & 	1.54$\pm$0.30 & 	2.09$\pm$0.20 & 	$-$2.76$\pm$0.14 \\
2MASS J01201051-4849461 & 	--- & 	31 & 	4797$\pm$150 & 	1.69$\pm$0.30 & 	1.08$\pm$0.20 & 	$-$2.17$\pm$0.14 \\
RAVE J012931.1-160046 & 	BPS CS 31082-0001, BD-16 251 & 	49 & 	4676$\pm$150 & 	1.01$\pm$0.30 & 	2.23$\pm$0.20 & 	$-$2.98$\pm$0.15 \\
RAVE J013807.0-195210 & 	--- & 	34 & 	4898$\pm$150 & 	2.48$\pm$0.30 & 	1.28$\pm$0.20 & 	$-$2.05$\pm$0.14 \\
RAVE J014254.2-503249 & 	--- & 	37 & 	5109$\pm$150 & 	2.69$\pm$0.30 & 	1.45$\pm$0.20 & 	$-$2.18$\pm$0.11 \\
RAVE J014908.0-491143 & 	HE 0147-4926, CD-49 506 & 	66 & 	4556$\pm$150 & 	0.72$\pm$0.30 & 	2.14$\pm$0.20 & 	$-$3.08$\pm$0.10 \\
\enddata
\tablecomments{This table is a stub; the full table is available in the electronic edition.}
\end{deluxetable*}

\begin{figure*}[ht]
    \centering
    \includegraphics[scale = 0.36]{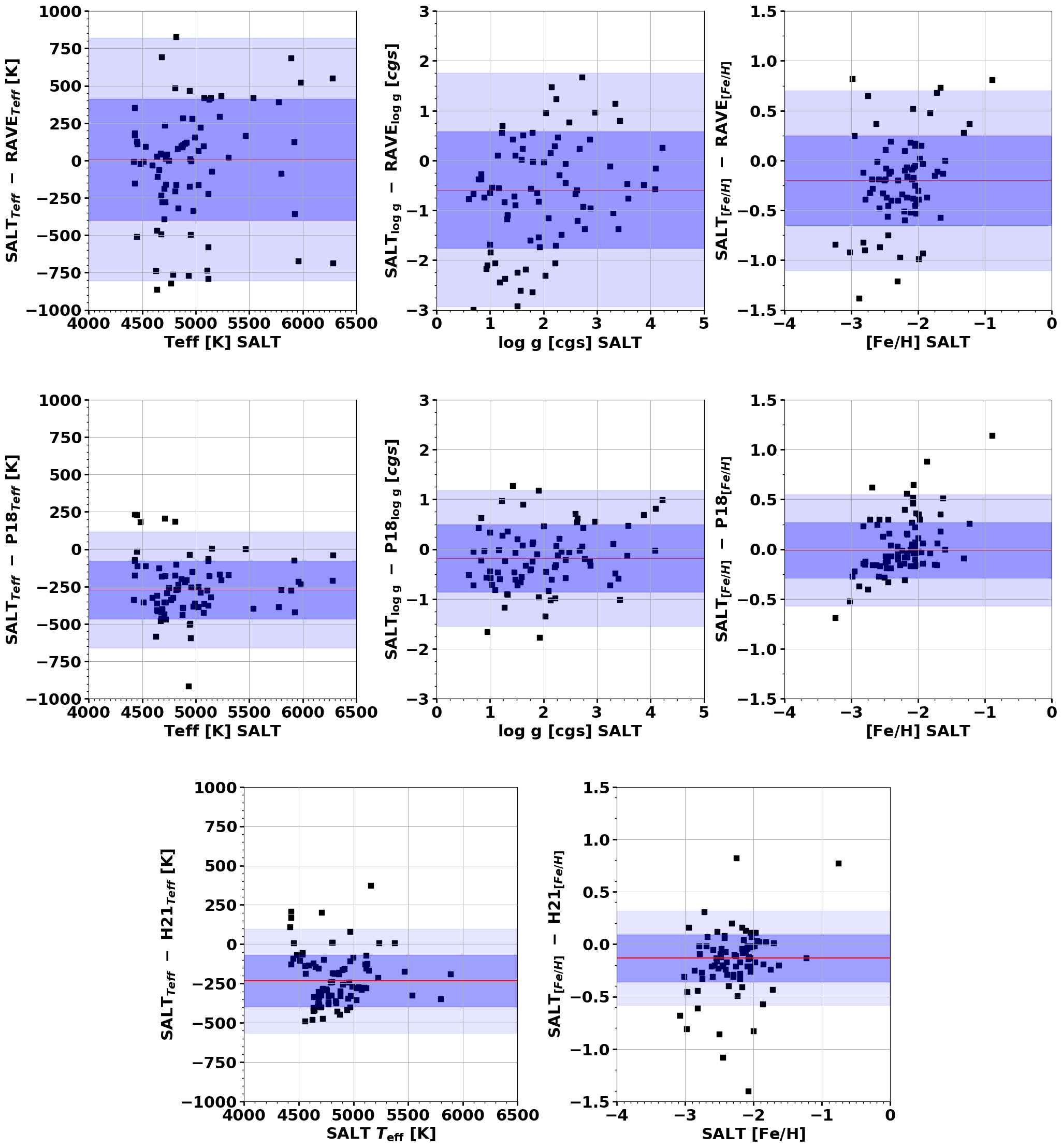}
    \caption{The residuals between our stellar parameters and RAVE DR5 (\citealt{Rave5}), \citet{Placco_2018}, and \citet{Huang_2021}. The top three plots show this sample compared to RAVE DR5 data. The middle plots show residuals between this sample and \citet{Placco_2018}. The bottom two plots show residuals of the differences between this sample and \citet{Huang_2021}.  The residuals are calculated as this sample's values minus the RAVE/Placco/Huang values. The red line shows the biweight location and the blue shaded regions show the first (1$\sigma$) and second (2$\sigma$) biweight scale ranges of the comparisons.}
    \label{fig:comp}
\end{figure*}

Stellar-parameter estimates were determined using Fe~I and Fe~II equivalent-width measurements. To determine $T_{\rm eff}$, Fe I abundances were measured from lines with a variety of excitation potentials, and the $T_{\rm eff}$ was chosen to minimize the residuals between iron abundances measured from the Fe~I lines. In order to determine the surface gravities, $\log g$, the residuals between the abundance measurements of Fe~I and Fe~II lines were minimized. Microturbulent velocities were found using Fe~I lines with varied reduced equivalent widths. The microturbulent velocity that gave the best agreement in iron abundance from the Fe~I lines of the various reduced equivalent widths was chosen. To find the stellar parameters that minimized all of these differences the \textit{solve} function within SMHR was used. The error in the stellar parameters are dominated by systematic error and thus we report this for simplicity. The resulting derived stellar parameters are shown in Figure \ref{fig:isochrone} (which includes 12 Gyr isochrones for a range of [Fe/H]) and listed in Table~\ref{tab:sp}, along with the S/N of the spectra.

\subsection{Comparisons with Previous Estimates}

We now consider comparisons between our derived stellar parameters and those provided by RAVE DR5 \citep{Rave5}, \citet{Placco_2018} (P18), and \citet{Huang_2021} (H21), shown in Figure \ref{fig:comp}.  For this exercise, we make use of the biweight estimators of location and scale, which are robust statistical estimators described in \citet{Beers_1990}. The biweight location is similar to a mean, but is less affected by outliers. The biweight scale describes the dispersion of the data, and is a resistant estimator similar to a standard deviation. We compare our $T_{\rm eff}$, surface gravity, and [Fe/H] to all three previous estimations of stellar parameters; no comparison for surface gravity is made with H21, because only luminosity classes were assigned in that paper.

The RAVE estimates of $T_{\rm eff}$ (top-left panel of Figure \ref{fig:comp}) have a negligible offset with respect to the SALT data (8\,K), however they exhibit large scatter (405\,K). Comparisons with the temperature results from P18 shown in the center-left panel exhibit an offset of $-271$\,K, but much reduced dispersion (194\,K), indicating significantly higher precision than obtained from the RAVE residuals. The offset for the comparisons with P18 can be partially explained by their use of photometric temperatures. As shown by \citet{Frebel2013}, photometric estimates of temperature tend to be higher than those based exclusively on spectroscopic data. The $T_{\rm eff}$ comparison with H21 shown in the bottom-left panel has an offset of $-$233\,K and a dispersion of 165\,K. The offset again is likely caused by the differences in the photometric and spectroscopic temperature estimates. The dispersion in the residuals for these estimates between our sample and H21 is lower than for the other two comparisons.

The surface-gravity residuals from RAVE (top-middle panel) with respect to the SALT analysis have an offset of $-0.59$\,dex, and a rather large dispersion (1.17\,{dex}). This large difference likely arises from the inherent degeneracy that comes from determining $T_{\rm eff}$, surface gravity, and [Fe/H] from a single feature (in this case the Ca triplet). RAVE appears to underestimate the surface gravity for stars with lower surface gravity and overestimate the surface gravity for stars with higher surface gravity, when compared to the SALT data. Comparison of the surface-gravity estimates between P18 (lower-middle panel) and the SALT stars yields a small offset, $-0.18$\,dex. The dispersion is also much smaller than for the RAVE data, 0.68\,dex, and does not exhibit any trend with respect to the SALT estimates.

The residuals in estimates of [Fe/H] from RAVE (top-right panel) with respect to the SALT data have a reasonably small offset ($-0.20$\,dex), with a dispersion of 0.45\,dex, but appear to exhibit a larger spread for stars with [Fe/H] $\leq -2.5$ compared to the SALT results. The [Fe/H] results from P18 (center-right panel) match-up  quite well with the SALT results. There is a negligible offset ($-0.01$\,dex) and relatively small dispersion (0.28\,dex). The comparisons to H21 exhibit a slight offset of $-$0.13\,dex with a small dispersion of 0.23\,dex. 

\section{Abundance Measurements} \label{sec:abun}
 
 The measurements of elemental abundances for the stars in this sample were accomplished using equivalent-width measurements and spectrum synthesis (for C and the n-capture elements). The line list used for equivalent-width measurements is provided in Table~\ref{tab:line_list}. This line list was generated using \texttt{Linemake} \citep{Placco_2021}. 
 
 \begin{deluxetable}{lllr}
\tablecaption{\label{tab:line_list}Line List for Equivalent-Width Analysis}
\tabletypesize{\scriptsize}
\tablehead{
\colhead{Ion}  & 
\colhead{$\lambda$} & 
\colhead{$\chi$} &
\colhead{log $gf$}
}
\startdata
Fe I  & 3963.1  & 3.281 & $-$0.68  \\
Fe I  & 3977.74 & 2.196 & $-$1.12  \\
Fe I  & 4001.66 & 2.174 & $-$1.90  \\
Fe I  & 4005.24 & 1.556 & $-$0.58  \\
Fe I  & 4032.63 & 1.484 & $-$2.38  \\
Fe I  & 4045.81 & 1.484 & 0.28   \\
Fe I  & 4058.22 & 3.209 & $-$1.18  \\
Fe I  & 4063.28 & 3.368 & $-$0.81  \\
Fe I  & 4063.59 & 1.556 & 0.06   \\
Fe I  & 4067.98 & 3.209 & $-$0.53  \\
\enddata
\tablecomments{This table is a stub; the full table is available in the electronic edition.}
\end{deluxetable}

The equivalent widths were first found for iron with measurements of Fe I and Fe II lines, and were used to find the stellar parameters, as described above. Then, equivalent widths were measured for Mg I, Si I, Ca I, Sc I, Sc II, Ti I, Ti II, Cr I, Cr II, Mn I, Co I, Ni I, Zn I, Sr II, and Ba II, when the spectrum permitted, using a Gaussian fit to the available features. The equivalent widths are then compared to interpolations of the ATLAS9 model atmospheres \citep{Castelli_2003}.

Stellar abundances are based on the mean value when multiple features for a given element were measured. Errors for these are taken as the standard deviation  of the measurements. 
 For species with only one measurement an error of 0.20\,dex is adopted. The results are listed in Tables \ref{tab:ab1} and \ref{tab:ab2}.
 
For the stellar synthesis of C, Ba, and Eu, MOOG (\citealt{Sneden_1973} and \citealt{Sobeck_2011}) was employed to carry out LTE analyses of the stellar spectra.  For the estimation of the stellar abundances, we use $\alpha$-enhanced ([$\alpha$/Fe] = +0.4) ATLAS9 model atmospheres \citep{Castelli_2003}. Line lists for each region of
interest are generated with \texttt{Linemake}\footnote{https://github.com/vmplacco/linemake} \citep{Placco_2021}. These line lists
include CH, C$_2$, and CN molecular lines \citep{Brooke_2013, Masseron_2014, Ram_2014, Sneden_2014}, as well as isotopic shift and hyperfine-structure information for Ba and Eu \citep{Lawler_2001, Gallagher_2010}. We use the Solar isotopic ratios in \citet{Asplund_2009} for n-capture elements with hyperfine-splitting effects. The C abundances were obtained by fitting the CH $G$-band at 4313\,{\AA}.

\subsection{Equivalent-Width Measurements}

There are eight strong Mg I lines in our available wavelength range; we were able to measure six of them for most stars. The feature at 3986\,{\AA} was measured for roughly three quarters of the stars. The feature at 4167\,{\AA} was blended with another feature for most stars, and so was measured successfully for only about 40\% of the stars.

There was one Si I line available for measurement (4103\,{\AA}), and we were able to measure it for about 70\,$\%$ of the sample. For the remaining stars, the strength of the C$_2$ feature at 4102\,{\AA} washed out the Si I line.

\begin{figure*}
    \centering
    \includegraphics[scale = 0.35]{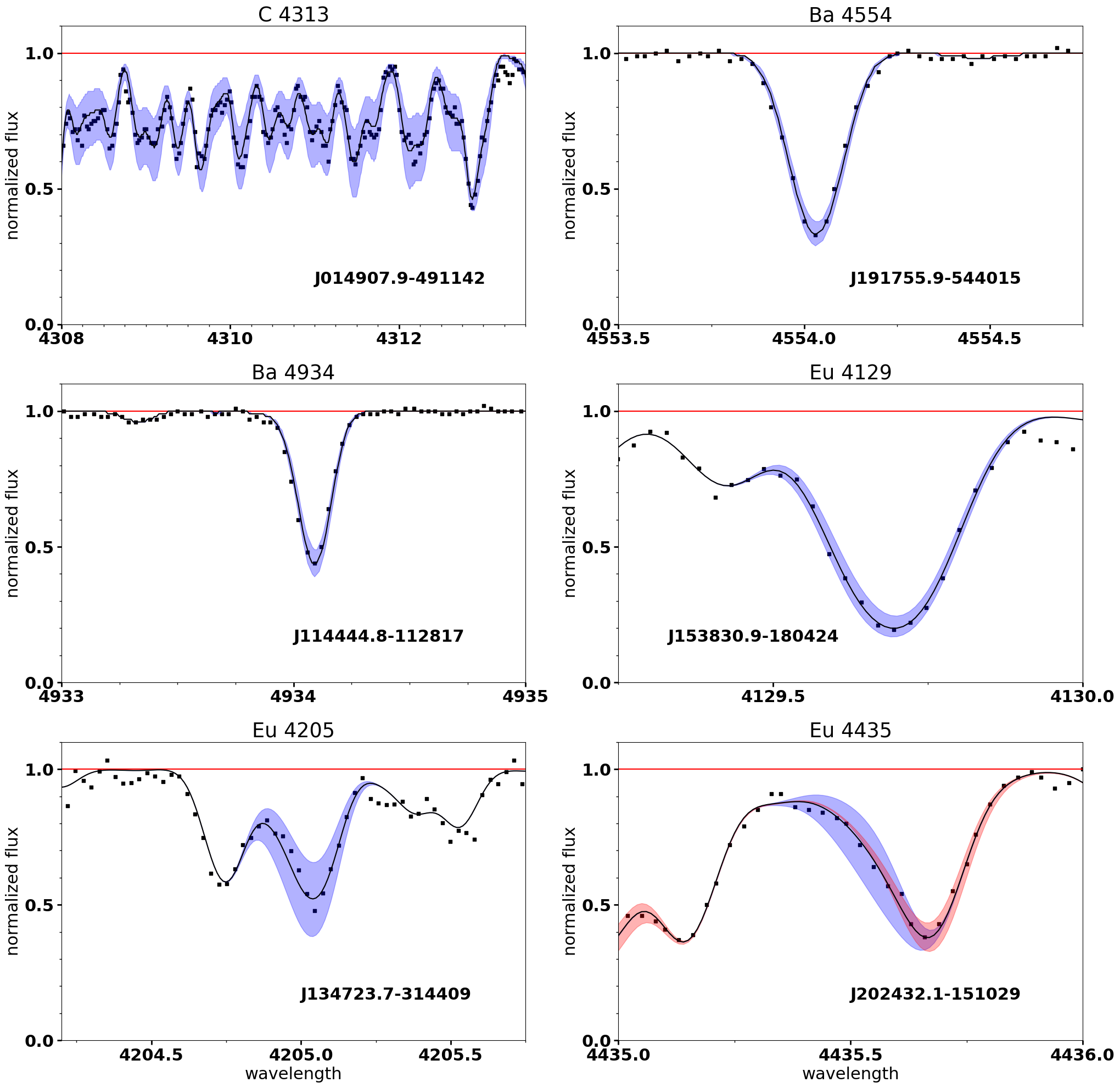}
    \caption{Examples of the synthesis measurements used in this study. The black dots represent the measured flux values and the blue lines represent the model spectra, with the blue region representing the error bars for the feature being measured. The black line represents the synthesis for the measured abundance. The red line represents the continuum. For the Eu 4435\,{\AA} feature, the error for Ca is shown by the red region since it is a blended feature. Each of these examples comes from a different star, shown in each panel. } 
    \label{fig:synth}
\end{figure*}

The Ca I lines, with the exception of the 4435\,{\AA} line, were available for measurement for most stars in the sample.  The 4435\,{\AA} line was blended with an adjacent feature, and thus was only measured for about half of the stars in the sample.

There was one Sc I line and nine Sc II lines in the wavelength coverage of this sample. Of these ten total lines, eight were able to be measured in most stars. The Sc I 4995\,{\AA} line and Sc II 5318\,{\AA} line were too weak in many stars for measurement. The number of Sc lines measured allowed for errors below 0.30\,dex for 80\% of the sample.

The abundance of Ti was estimated with both Ti I and Ti II lines, which have many lines available for measurement in this sample. On average, we were able measure 21 Ti I lines, and 34 Ti II lines in each star. The resulting error for the Ti abundance estimate was, on average, 0.17\,dex. There were, however, a few lines that were rarely used; these lines are Ti I at 4008\,{\AA}, and Ti II at 4398\,{\AA}, 4409\,{\AA}, 4636\,{\AA}, and 5268\,{\AA}. 

Both Cr I and Cr II have measurable features in the observed wavelength range. Between the two species an average of 25 Cr lines are measured per star, with an average dispersion of the derived abundance estimates of 0.16\,dex. 

Manganese had few lines available, but most were readily measurable; we measured four lines for most stars. We were able to measure the weak lines at 4034\,{\AA} and 4041\,{\AA} for about 75\% of the sample. The situation for the Co I lines is very similar to Mn; three lines were measured on average. We were able to measure the 3995\,{\AA} and 4110\,{\AA} lines for approximately 90\% of the sample, the 4020\,{\AA} line was measured for 75\% of the stars, and the 4118\,{\AA} line was measured for half the stars.

Nickel has many lines in the spectral region observed,  with most stars having more than ten lines measured. The Ni abundance estimates exhibit good agreement, with over 90\% of stars having an error in Ni no greater than 0.20\,dex. The lines at 4605\,{\AA}, 4686\,{\AA}, and 5155\,{\AA} were able to be measured for about two-thirds of the stars, despite their weakness.

The last three elements with abundance estimates measured with equivalent widths, Zn, Sr, and Ba, each had two features that could be measured -- for Zn, the lines at 4722\,{\AA} and 4810\,{\AA}, for Sr, the lines at  4077\,{\AA} and 4215\,{\AA}, and for Ba, the lines at 4554\,{\AA} and 4934\,{\AA}. These features were measured for about 90\,$\%$ of the stars, except for the Sr II line at 4077\,\AA. This line was only possible to measure in about 30\,$\%$ of the stars, due to blending.

\subsection{Synthesis Measurements}

Synthesis measurements were made for six spectral features---three Eu features, two Ba features, and one C feature. Figure \ref{fig:synth} shows examples of all of these syntheses. The errors for these abundance estimates are 0.20\,dex for Ba and Eu and 0.30\,dex for C, based on the systematic errors arising from errors in stellar parameters as described in \citet{Holmbeck_2020} for spectra of similar resolution, S/N, and stellar parameters. The $1\sigma$ errors in the fits are shown as the shaded blue regions in Figure \ref{fig:synth}. These are all significantly smaller than the expected systematic errors, so we simply adopt the systematic errors in our estimated abundances. To account for isotopic differences in these syntheses we considered the following isotopes:  $^{12}$C, $^{13}$C, $^{134}$Ba, $^{135}$Ba, $^{136}$Ba, $^{137}$Ba, $^{138}$Ba, $^{151}$Eu, and $^{153}$Eu.

The molecular C feature at 4313\,{\AA} extends across a wide wavelength region, as shown in Figure \ref{fig:synth}.  Carbon abundances were corrected for depletion on the giant branch using the method prescribed in \citet{Placco_2014}.

The two Ba features that were used for synthesis are the same ones used for equivalent-width measurements. Both features are relatively easy to measure due to the lack of other nearby features, the S/N at their locations, and the strength of the features. Each feature was measured for over 80\,$\%$ of the stars; blending prevented measurement for the remaining 20\,$\%$ of the stars. 

The Eu feature at 4129\,{\AA} has lower S/N than the other synthesized features. This feature was able to be measured through synthesis for about half the stars. The  Eu feature at 4205\,{\AA} was measured for roughly 60\,$\%$ of the sample stars. The nearby Mn II feature and CH isotopic features affect the measurement of the synthesis in the stars for which Eu wasn't measured at this location. The Eu feature at 4435\,{\AA} is blended with a Ca I feature, so accurate measurement of Ca is important for the measurement of europium with this feature. The blend with Ca was handled sufficiently well for just over half of the sample, and thus allowed a measurement of Eu based on this feature.

Where abundance analysis results were available for both the equivalent-width and synthesis methods, we adopt the mean of these determinations.  The full set of results are provided in Tables \ref{tab:ab1}, \ref{tab:ab2}, and \ref{tab:ab3}. 

\section{Results} \label{sec:result}

\begin{deluxetable*}{cccccc}
\tablecaption{\label{tab:ab1} Abundances of Alpha Elements}
\tabletypesize{\scriptsize}
\tablehead{
\colhead{NAME}  & 
\colhead{[Mg/Fe]} & 
\colhead{[Si/Fe]} &  
\colhead{[Ca/Fe]} &
\colhead{[Sc/Fe]} & 
\colhead{[Ti/Fe]} \\
&
\colhead{log$\epsilon$(Mg)} &
\colhead{log$\epsilon$(Si)} &
\colhead{log$\epsilon$(Ca)} &
\colhead{log$\epsilon$(Sc)} &
\colhead{log$\epsilon$(Ti)} 
}
\startdata
RAVE J000438.4-540134 & 	 +0.39$\pm$0.30 [4] & 	 +1.06$\pm$0.20 [1] & 	 +0.58$\pm$0.16 [7] & 	 +0.38$\pm$0.34 [4] & 	 +0.43$\pm$0.24 [53] \\	 &   5.61 &   6.19 &   4.54 &   1.15 &   3.00 \\
RAVE J001236.5-181631 & 	 +0.51$\pm$0.24 [6] & 	 +0.67$\pm$0.20 [1] & 	 +0.26$\pm$0.13 [10] & 	 +0.01$\pm$0.13 [4] & 	 +0.25$\pm$0.17 [69] \\	 &   5.70 &   5.77 &   4.19 &   0.75 &   2.79 \\
RAVE J003946.9-684957 & 	 +0.55$\pm$0.10 [4] & 	 ... & 	 +0.60$\pm$0.31 [6] & 	  $-$0.07$\pm$0.10 [2] & 	 +0.32$\pm$0.14 [41] \\	 &   5.53 &  ... &   4.32 &   0.46 &   2.65 \\
RAVE J010839.6-285701 & 	 +0.41$\pm$0.10 [3] & 	 ... & 	 +0.32$\pm$0.11 [2] & 	 +0.01$\pm$0.28 [2] & 	 +0.14$\pm$0.17 [19] \\	 &   5.14 &  ... &   3.79 &   0.29 &   2.22 \\
2MASS J01162885-5343408 & 	 +0.45$\pm$0.13 [8] & 	 +1.17$\pm$0.20 [1] & 	 +0.36$\pm$0.12 [8] & 	 +0.04$\pm$0.24 [4] & 	 +0.33$\pm$0.17 [62] \\	 &   5.29 &   5.92 &   3.94 &   0.43 &   2.52 \\
2MASS J01201051-4849461 & 	 +0.26$\pm$0.10 [3] & 	 ... & 	 +0.46$\pm$0.20 [6] & 	  $-$0.06$\pm$0.10 [3] & 	 +0.30$\pm$0.16 [47] \\	 &   5.69 &  ... &   4.63 &   0.92 &   3.08 \\
RAVE J012931.1-160046 & 	 +0.46$\pm$0.17 [5] & 	 ... & 	 +0.54$\pm$0.10 [7] & 	  $-$0.03$\pm$0.10 [7] & 	 +0.28$\pm$0.20 [53] \\	 &   5.08 &  ... &   3.90 &   0.14 &   2.25 \\
RAVE J013807.0-195210 & 	 +0.21$\pm$0.38 [5] & 	 +1.16$\pm$0.20 [1] & 	 +0.29$\pm$0.20 [6] & 	 +0.23$\pm$0.10 [4] & 	 +0.53$\pm$0.25 [61] \\	 &   5.76 &   6.62 &   4.58 &   1.33 &   3.43 \\
RAVE J014254.2-503249 & 	 +0.33$\pm$0.13 [6] & 	 ... & 	 +0.39$\pm$0.16 [9] & 	 +0.20$\pm$0.23 [7] & 	 +0.35$\pm$0.15 [66] \\	 &   5.75 &  ... &   4.55 &   1.17 &   3.12 \\
RAVE J014908.0-491143 & 	 +0.63$\pm$0.17 [5] & 	 ... & 	 +0.61$\pm$0.13 [8] & 	 +0.01$\pm$0.13 [7] & 	 +0.26$\pm$0.12 [52] \\	 &   5.15 &  ... &   3.87 &   0.08 &   2.13 \\
\enddata
\tablecomments{This table is a stub; the full table is available in the electronic edition. The ``[\#]" denote the number of lines used for measurement.}

\end{deluxetable*}

\begin{deluxetable*}{ccccccc}
\tablecaption{\label{tab:ab2} Abundances of Iron-Peak Elements}
\tabletypesize{\scriptsize}
\tablehead{
\colhead{NAME}  & 
\colhead{[Cr/Fe]} & 
\colhead{[Mn/Fe]} &  
\colhead{[Co/Fe]} &
\colhead{[Fe/H]} &
\colhead{[Ni/Fe]} & 
\colhead{[Zn/Fe]} \\
&
\colhead{log$\epsilon$(Cr)} &
\colhead{log$\epsilon$(Mn)} &
\colhead{log$\epsilon$(Co)} &
\colhead{log$\epsilon$(Fe)} &
\colhead{log$\epsilon$(Ni)} &
\colhead{log$\epsilon$(Zn)} 
}
\startdata
RAVE J000438.4-540134 & 	  $-$0.09$\pm$0.09 [11]  & 	  $-$0.38$\pm$0.10 [3]  & 	 ...  & 	  $-$2.38$\pm$0.10 [74]  & 	 +0.05$\pm$0.12 [8]  & 	 +0.21$\pm$0.10 [2]  \\	 &   3.17 &   2.67 &  ... &   5.12 &   3.89 &   2.39 \\
RAVE J001236.5-181631 & 	  $-$0.14$\pm$0.14 [20]  & 	 +0.06$\pm$0.35 [7]  & 	 +0.62$\pm$0.31 [4]  & 	  $-$2.41$\pm$0.09 [106]  & 	 +0.16$\pm$0.10 [14]  & 	 +0.12$\pm$0.10 [2]  \\	 &   3.09 &   3.08 &   3.2 &   5.09 &   3.97 &   2.27 \\
RAVE J003946.9-684957 & 	  $-$0.17$\pm$0.15 [8]  & 	  $-$0.47$\pm$0.20 [1]  & 	 +0.52$\pm$0.2 [1]  & 	  $-$2.62$\pm$0.10 [79]  & 	 +0.09$\pm$0.10 [5]  & 	 +0.27$\pm$0.11 [2]  \\	 &   2.85 &   2.34 &   2.89 &   4.88 &   3.69 &   2.21 \\
RAVE J010839.6-285701 & 	  $-$0.31$\pm$0.24 [5]  & 	 ...  & 	 ...  & 	  $-$2.87$\pm$0.14 [77]  & 	 +0.04$\pm$0.16 [3]  & 	 +0.03$\pm$0.20 [1]  \\	 &   2.46 &  ... &  ... &   4.63 &   3.39 &   1.72 \\
2MASS J01162885-5343408 & 	  $-$0.50$\pm$0.22 [16]  & 	  $-$0.30$\pm$0.53 [6]  & 	 +0.03$\pm$0.27 [2]  & 	  $-$2.76$\pm$0.14 [104]  & 	 +0.09$\pm$0.19 [7]  & 	 +0.06$\pm$0.20 [1]  \\	 &   2.38 &   2.37 &   2.77 &   4.74 &   3.55 &   1.86 \\
2MASS J01201051-4849461 & 	  $-$0.08$\pm$0.12 [8]  & 	  $-$0.38$\pm$0.10 [3]  & 	 ...  & 	  $-$2.17$\pm$0.14 [103]  & 	 +0.07$\pm$0.10 [9]  & 	 +0.19$\pm$0.12 [2]  \\	 &   3.39 &   2.88 &  ... &   5.33 &   4.12 &   2.58 \\
RAVE J012931.1-160046 & 	  $-$0.19$\pm$0.18 [11]  & 	  $-$0.31$\pm$0.20 [1]  & 	 +0.52$\pm$0.2 [1]  & 	  $-$2.98$\pm$0.15 [112]  & 	 +0.16$\pm$0.12 [6]  & 	 +0.14$\pm$0.26 [2]  \\	 &   2.47 &   2.14 &   2.53 &   4.52 &   3.40 &   1.72 \\
RAVE J013807.0-195210 & 	  $-$0.20$\pm$0.17 [12]  & 	  $-$0.23$\pm$0.23 [4]  & 	  $-$0.36$\pm$0.19 [2]  & 	  $-$2.05$\pm$0.14 [87]  & 	 +0.10$\pm$0.21 [10]  & 	 +0.27$\pm$0.10 [2]  \\	 &   3.39 &   3.15 &   3.18 &   5.45 &   4.27 &   2.78 \\
RAVE J014254.2-503249 & 	  $-$0.09$\pm$0.15 [22]  & 	  $-$0.39$\pm$0.10 [3]  & 	 +1.1$\pm$0.2 [1]  & 	  $-$2.18$\pm$0.11 [108]  & 	 +0.04$\pm$0.11 [15]  & 	 +0.25$\pm$0.12 [2]  \\	 &   3.37 &   2.86 &   3.91 &   5.32 &   4.08 &   2.63 \\
RAVE J014908.0-491143 & 	  $-$0.17$\pm$0.13 [13]  & 	  $-$0.24$\pm$0.18 [5]  & 	 +0.94$\pm$0.1 [2]  & 	  $-$3.08$\pm$0.10 [105]  & 	 +0.18$\pm$0.14 [10]  & 	 +0.29$\pm$0.20 [1]  \\	 &   2.39 &   2.11 &   2.85 &   4.42 &   3.32 &   1.77 \\
\enddata
\tablecomments{This table is a stub; the full table is available in the electronic edition. The ``[\#]" denote the number of lines used for measurement.}

\end{deluxetable*}

\begin{deluxetable*}{ccccccccc}
\tablecaption{\label{tab:ab3}Classifications, Carbon, and Neutron-capture Elements}
\tabletypesize{\scriptsize}
\tablehead{
\colhead{NAME}  &
\colhead{Class}  &
\colhead{[C/Fe]} & 
\colhead{[C/Fe]$_{c}$} &
\colhead{$A$(C)$_{c}$} &
\colhead{[Sr/Fe]} & 
\colhead{[Ba/Fe] (syn)} & 
\colhead{[Eu/Fe] (syn)} &  
\colhead{[Ba/Eu]} \\
 &
 &
 &
 &
 &
 \colhead{log$\epsilon$(Sr)} &
 \colhead{log$\epsilon$(Ba)} &
 \colhead{log$\epsilon$(Eu)} &
 }
\startdata
RAVE J000438.4-540134 & 	$r$-I & 	 +0.06$\pm$0.30 [1] & 	 +0.55$\pm$0.30 & 	6.60 & 	 +0.35$\pm$0.20 [1] &	 +0.16$\pm$0.20 [2]  &	 +0.61$\pm$0.20 [2 ] &	  $-$0.45$\pm$0.28 \\	 &  &  &  &  &   0.84 &  $-$0.04 &  $-$1.25 &  \\
RAVE J001236.5-181631 & 	... & 	  $-$0.01$\pm$0.30 [1] & 	 +0.00$\pm$0.30 & 	6.02 & 	  $-$0.14$\pm$0.10 [2] &	  $-$0.60$\pm$0.20 [2]  &	 +0.17$\pm$0.20 [2 ] &	  $-$0.76$\pm$0.28 \\	 &  &  &  &  &   0.32 &  $-$0.83 &  $-$1.72 &  \\
RAVE J003946.9-684957 & 	CEMP-no & 	 +0.10$\pm$0.30 [1] & 	 +0.73$\pm$0.30 & 	6.54 & 	 +0.43$\pm$0.20 [1] &	  $-$0.10$\pm$0.20 [1]  &	 +0.28$\pm$0.20 [3 ] &	  $-$0.38$\pm$0.28 \\	 &  &  &  &  &   0.68 &  $-$0.54 &  $-$1.82 &  \\
RAVE J010839.6-285701 & 	CEMP-$r$ & 	 +0.10$\pm$0.30 [1] & 	 +0.76$\pm$0.30 & 	6.32 & 	 +0.41$\pm$0.20 [1] &	  $-$0.52$\pm$0.20 [2]  &	 +0.45$\pm$0.20 [1 ] &	  $-$0.97$\pm$0.28 \\	 &  &  &  &  &   0.41 &  $-$1.21 &  $-$1.90 &  \\
2MASS J01162885-5343408 & 	... & 	  $-$0.02$\pm$0.30 [1] & 	 +0.39$\pm$0.30 & 	6.06 & 	  $-$0.19$\pm$0.20 [1] &	 +0.17$\pm$0.20 [1]  &	 ... &	 ... \\	 &  &  &  &  &  $-$0.08 &  $-$0.41 &  ... &  \\
2MASS J01201051-4849461 & 	... & 	  $-$0.12$\pm$0.30 [1] & 	 +0.20$\pm$0.30 & 	6.46 & 	 +0.03$\pm$0.20 [1] &	 +0.08$\pm$0.20 [2]  &	 +0.03$\pm$0.20 [1 ] &	 +0.06$\pm$0.28 \\	 &  &  &  &  &   0.73 &   0.09 &  $-$1.62 &  \\
RAVE J012931.1-160046 & 	CEMP-$r$ & 	 +0.26$\pm$0.30 [1] & 	 +0.95$\pm$0.30 & 	6.40 & 	 +0.45$\pm$0.20 [1] &	 +1.11$\pm$0.20 [2]  &	 +1.68$\pm$0.20 [3 ] &	  $-$0.57$\pm$0.28 \\	 &  &  &  &  &   0.34 &   0.31 &  $-$0.78 &  \\
RAVE J013807.0-195210 & 	$r$-I & 	  $-$0.09$\pm$0.30 [1] & 	  $-$0.08$\pm$0.30 & 	6.30 & 	 +0.05$\pm$0.20 [1] &	 +0.30$\pm$0.20 [2]  &	 +0.46$\pm$0.20 [1 ] &	  $-$0.16$\pm$0.28 \\	 &  &  &  &  &   0.87 &   0.43 &  $-$1.07 &  \\
RAVE J014254.2-503249 & 	$r$-I & 	 +0.25$\pm$0.30 [1] & 	 +0.26$\pm$0.30 & 	6.51 & 	 +0.08$\pm$0.20 [1] &	 +0.04$\pm$0.20 [2]  &	 +0.40$\pm$0.20 [2 ] &	  $-$0.36$\pm$0.28 \\	 &  &  &  &  &   0.77 &   0.04 &  $-$1.26 &  \\
RAVE J014908.0-491143 & 	CEMP-$r$ & 	 +0.07$\pm$0.30 [1] & 	 +0.81$\pm$0.30 & 	6.16 & 	  $-$0.01$\pm$0.20 [1] &	  $-$0.39$\pm$0.20 [2]  &	 +0.35$\pm$0.20 [3 ] &	  $-$0.74$\pm$0.28 \\	 &  &  &  &  &  $-$0.22 &  $-$1.29 &  $-$2.21 &  \\
\enddata
\tablecomments{This table is a stub; the full table is available in the electronic edition. The ``[\#]" denote the number of lines used for measurement. Both [C/Fe]$_{c}$ and $A$(C)$_{c}$ are corrected for evolutionary effects.}

\end{deluxetable*}

\subsection{Classifications}

\begin{figure}
    \centering
    \includegraphics[scale = 0.32]{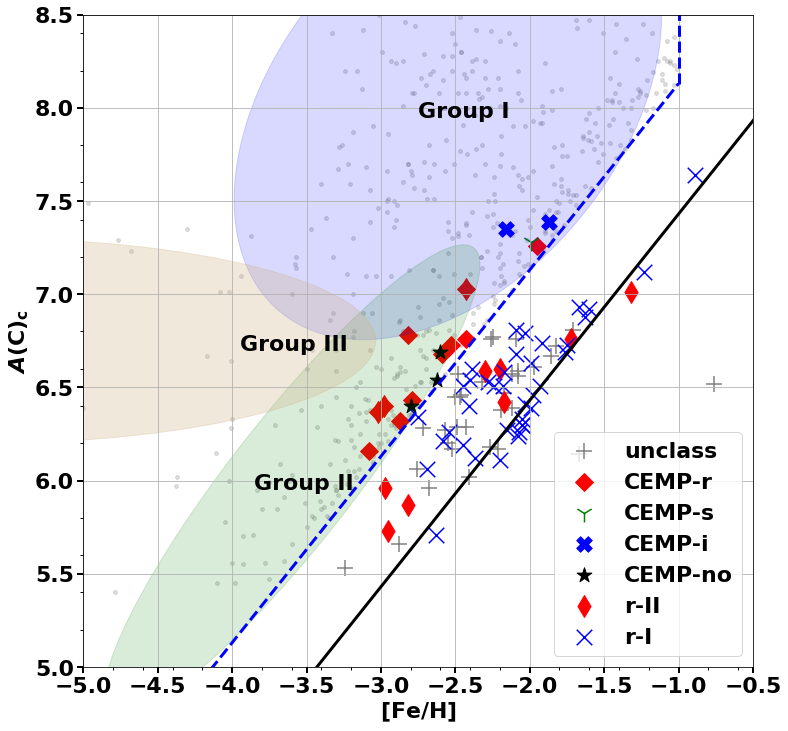}
    \caption{Yoon-Beers diagram of $A$(C)$_c$ vs. [Fe/H] for the sample stars with available carbon measurements, corrected for evolutionary effects. The blue dashed line shows the adopted criterion for CEMP stars, [C/Fe] $> +0.7$.  The black solid line corresponds to [C/Fe] = 0.  The sub-classifications assigned in this work are indicated as shown in the legend. The different morphological groups of CEMP stars defined by \citet{Yoon_2016} are shown by the colored ellipses. Literature CEMP stars from the SAGA database \citep{Saga} are shown by the gray dots.}
    \label{fig:YB}
\end{figure}

The abundance measurements are used to classify stars using the categories described in Table \ref{tab:classes}. \citet{Goswami_2021} and others have suggested that using [La/Eu] to determine $i$-process enrichment is more appropriate than [Ba/Eu], but in the absence of La we used [Ba/Eu]. The carbon and n-capture elemental abundances are listed in Table \ref{tab:ab3}. It should be noted that we report both evolutionary corrected values and uncorrected values of carbon in this table. A substantial fraction of the stars in this sample exhibit carbon enhancement, with 27 stars (24\,\% of the sample) having [C/Fe] $>$ +0.5. This sample contains 17 CEMP stars (15\,\% of the sample). The stars in the sample with measured carbon abundances are shown in the Yoon-Beers $A$(C)$_c$ vs. [Fe/H] diagram in Figure \ref{fig:YB}. Of the 17 CEMP stars, 11 are identified as CEMP-$r$ stars (8 newly identified), 1 as a CEMP-$s$ star (newly identified), 2 are identified as possible CEMP-$i$ stars (1 newly identified), and 3 are identified as CEMP-no stars (all newly identified). Among the stars with $r$-process enhancements, we found 11 $r$-II stars (8 newly identified, 38 $r$-I stars (31 newly identified), and 1 newly identified limited-$r$ star .

\subsection{Abundances}

\begin{figure*}
    \centering
    \includegraphics[scale = 0.35]{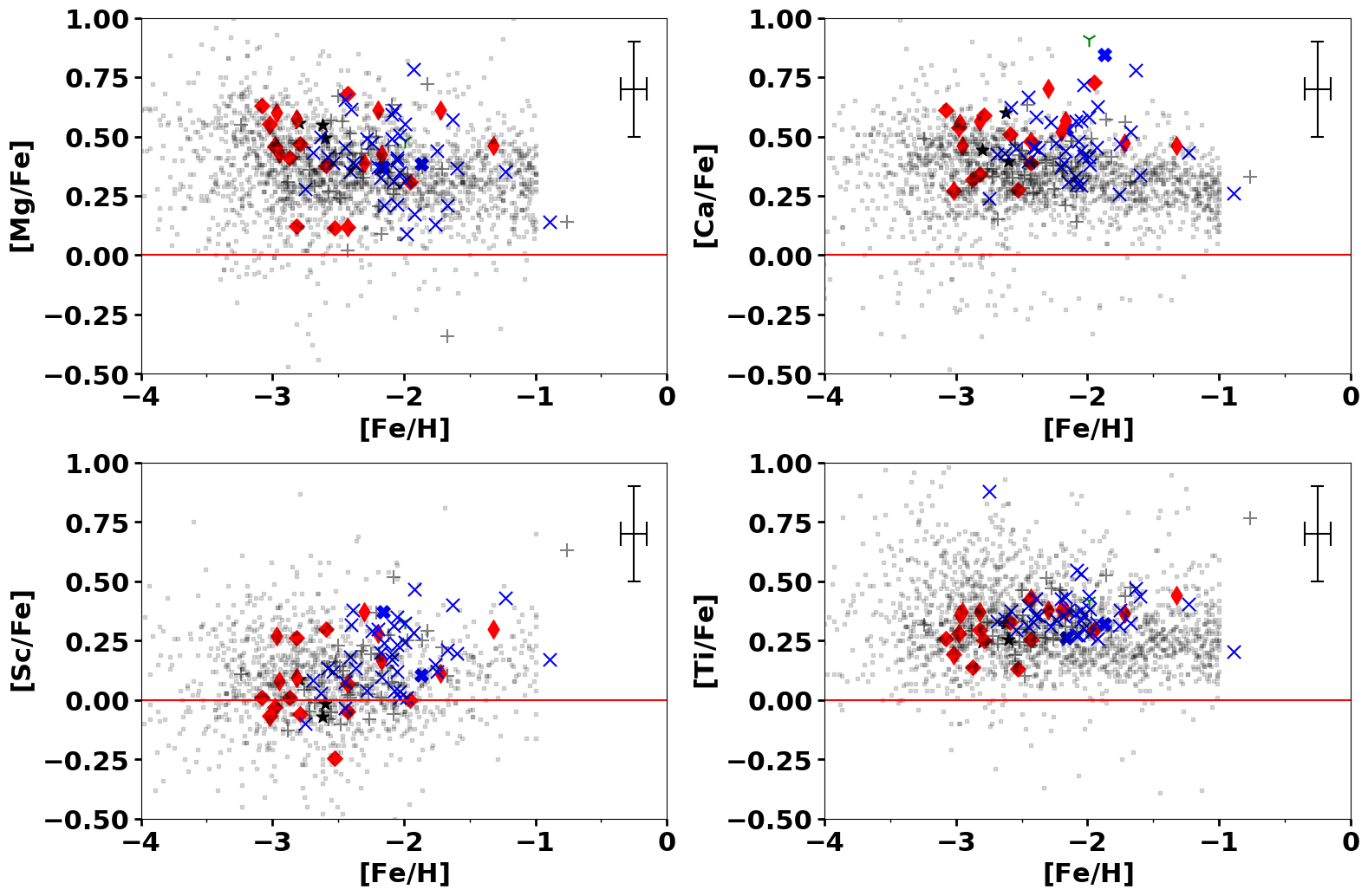}
    \caption{Abundances of the $\alpha$-capture elements for stars in our sample, sub-classified as shown in the legend of Figure \ref{fig:YB}. These stars are plotted along with MP stars taken from the SAGA database \citep{Saga} in light gray. An error bar of 0.1\,dex is shown for Fe and 0.20\,dex for the $\alpha$-capture elements.} 
    \label{fig:alpha}
\end{figure*}

\begin{figure*}
    \centering
    \includegraphics[scale = 0.35]{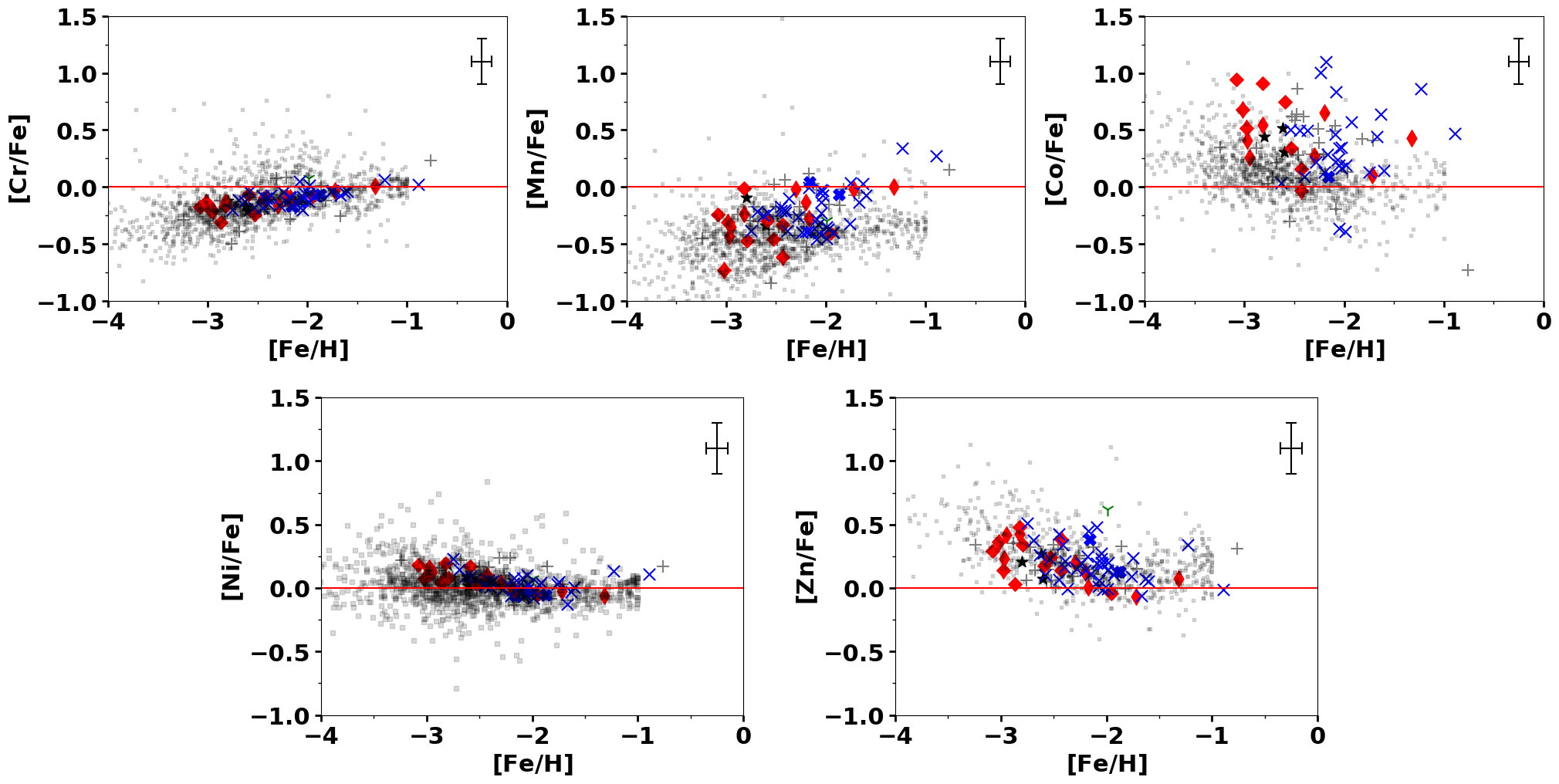}
    \caption{Abundances of the Fe-peak elements for stars in our sample, sub-classified as shown in the legend of Figure \ref{fig:YB}. These stars are plotted along with MP stars taken from the SAGA database \citep{Saga} in light gray. An error bar of 0.1\,dex is shown for Fe and 0.20\,dex for the Fe-peak elements.}
     \label{fig:fepeak}
\end{figure*}

\begin{figure*}
    \centering
    \includegraphics[scale = 0.35]{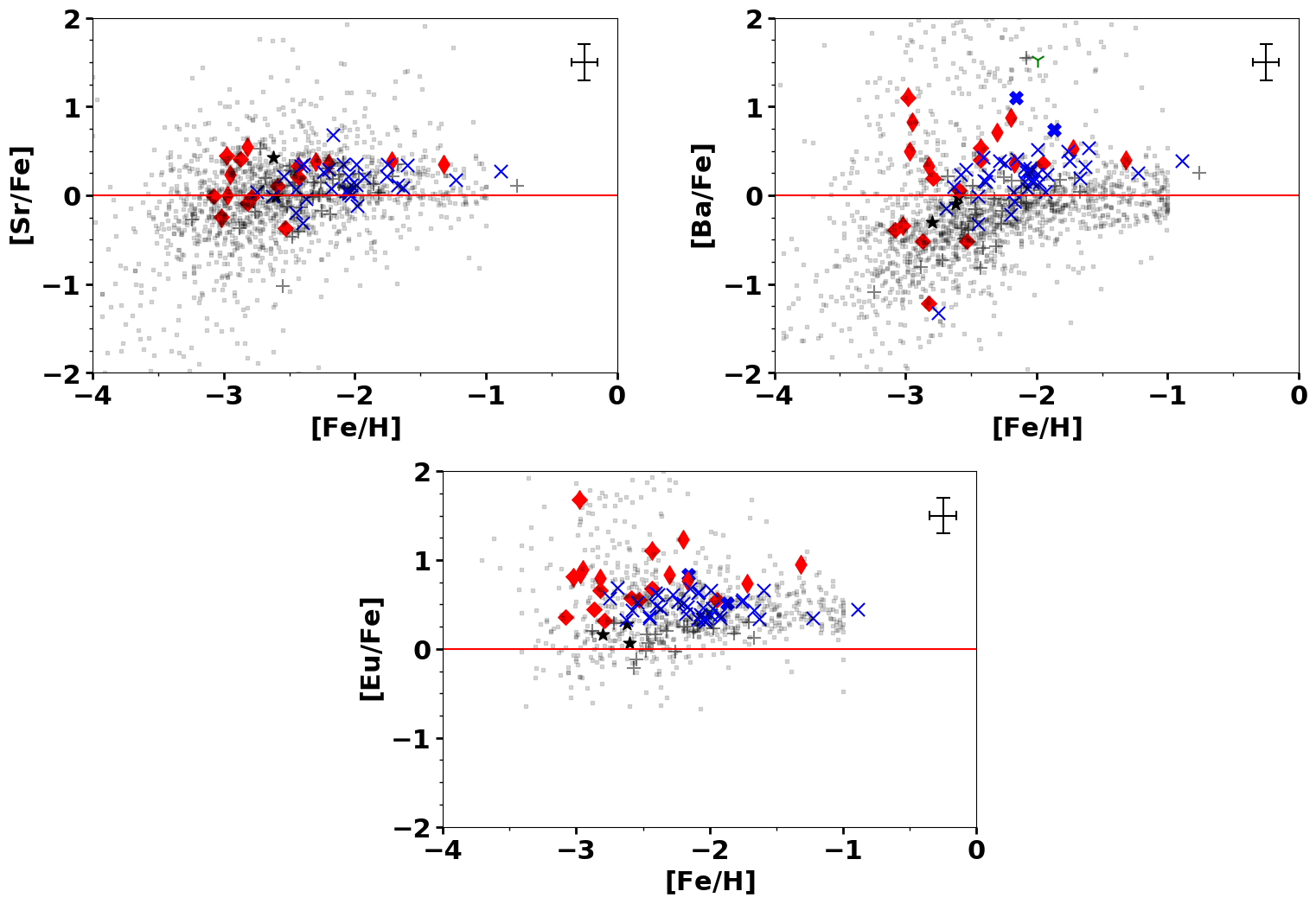}
    \caption{Abundances of n-capture elements for star in our sample, sub-classified as shown in the legend of Figure \ref{fig:YB}. These stars are plotted along with MP stars taken from the SAGA database \citep{Saga} in light gray. An error bar of 0.1\,dex is shown for Fe and 0.20\,dex for the n-capture elements.}
     \label{fig:ncapture}
\end{figure*}

Figures \ref{fig:alpha}, \ref{fig:fepeak}, and \ref{fig:ncapture} show the abundances measured in this work plotted over literature abundance values for MP stars taken from the SAGA database \citep{Saga}. With the exception of Co for our more metal-rich stars, the elements measured in our sample exhibit values and trends that are similar to those found for previously studied MP stars. \citet{Kobayashi_2009} discusses the differences in formation rates of Mn and Zn in type Ia supernova with varying metallicity. This difference in the formation rates is believed to be the cause for the abundances trends seen for Mn and Zn in Figure \ref{fig:fepeak}.

\begin{deluxetable*}{llllccccccccc}
\tablecaption{\label{tab:prev_params}Comparison with Previous High-Resolution Spectroscopic Results}
\tabletypesize{\scriptsize}
\tablehead{
\colhead{Name}  &
\colhead{Lit Name} &
\colhead{Lit.} & 
\colhead{S/N} &
\multicolumn{5}{c}{Difference From Literature}\\
\cline{5-8}
\colhead{} &
\colhead{} &
\colhead{} &
\colhead{} &
\colhead{$\Delta T_{\rm eff}$ \, (K)} & 
\colhead{$\Delta$ log $g$\,(dex)} & 
\colhead{$\Delta \xi$\,(km s$^{-1}$)} & 
\colhead{$\Delta$ [Fe/H]\,(dex)}
}
\startdata
RAVE J010839.6-285701 & 	3 & 	SMSS J010839.58-285701.5 & 	27 & 	$-$12 & 	$-$0.38 & 	$-$0.29 & 	+0.15 \\
RAVE J012931.1-160046 & 	1 &	BPS CS 31082-0001, BD-16 251 & 	49 & 	$-$200 & 	$-$0.79 & 	+0.10 & 	$-$0.17 \\
RAVE J014254.2-503249 & 	1 &	 & 	37 & 	$-$23 & 	+0.04 & 	$-$0.10 & 	$-$0.09 \\
RAVE J014908.0-491143 & 	1 &	HE 0147-4926, CD-49 506 & 	66 & 	$-$126 & 	$-$0.21 & 	+0.19 & 	$-$0.14 \\
RAVE J021110.2-191124 & 	2 &	 & 	39 & 	+248 & 	+0.82 & 	+0.13 & 	+0.10 \\
RAVE J114444.8-112817 & 	2 &	BD-10 3332 & 	70 & 	$-$18 & 	$-$0.05 & 	+0.10 & 	$-$0.21 \\
RAVE J133352.8-262354 & 	2 &	 & 	54 & 	$-$53 & 	$-$0.18 & 	$-$0.01 & 	$-$0.15 \\
RAVE J140335.4-333526 & 	1 &	CD-33 9514 & 	27 & 	$-$13 & 	+0.01 & 	+0.14 & 	$-$0.04 \\
RAVE J155829.6-122434 & 	1 &	 & 	63 & 	$-$288 & 	$-$0.52 & 	$-$0.14 & 	$-$0.28 \\
RAVE J160245.0-152102 & 	1 &	BD-14 4336 & 	81 & 	$-$214 & 	$-$0.54 & 	$-$0.40 & 	$-$0.25 \\
RAVE J180242.3-440443 & 	1 &	 & 	28 & 	$-$252 & 	$-$0.91 & 	$-$0.39 & 	$-$0.33 \\
RAVE J183623.2-642812 & 	1 &	 & 	28 & 	+6 & 	$-$0.01 & 	$-$0.12 & 	+0.03 \\
RAVE J192325.2-583341 & 	1 &	 & 	48 & 	$-$206 & 	$-$0.32 & 	$-$0.22 & 	$-$0.33 \\
RAVE J203622.6-071420 & 	2 &	 & 	81 & 	+144 & 	+0.76 & 	+0.71 & 	$-$0.18 \\
RAVE J205149.7-615801 & 	1 &	 & 	66 & 	$-$168 & 	$-$0.26 & 	$-$0.51 & 	$-$0.16 \\
RAVE J205609.1-133118 & 	1 &	BD-14 5890 , HD 358059 & 	93 & 	+24 & 	+0.06 & 	$-$0.01 & 	$-$0.03 \\
RAVE J210642.9-682827 & 	1 &	 & 	58 & 	$-$244 & 	$-$0.77 & 	$-$1.25 & 	$-$0.19 \\
RAVE J211525.5-150331 & 	1 &	 & 	21 & 	+85 & 	+0.44 & 	+0.30 & 	$-$0.02 \\
RAVE J224927.6-223829 & 	1 &	 & 	66 & 	$-$338 & 	$-$0.67 & 	$-$0.48 & 	$-$0.21 \\
RAVE J225319.9-224856 & 	1 &	CD-23 17654 & 	21 & 	$-$211 & 	$-$0.43 & 	+0.48 & 	$-$0.29 \\
RAVE J230222.9-683323 & 	1 &	 & 	50 & 	+312 & 	+0.74 & 	$-$0.26 & 	+0.38 \\
RAVE J231300.0-450707 & 	1 &	 & 	62 & 	+192 & 	+0.50 & 	+0.50 & 	+0.11 \\
 & 	 & 	 & 	 & 	$\mu$ = $-$63 & 	$\mu$ = +0.18 & 	$\mu$ = +0.04 & 	$\mu$ = +0.13 \\
 & 	 & 	 & 	 & 	$\sigma$ = 183 & 	$\sigma$ = 0.53 & 	$\sigma$ = 0.38 & 	$\sigma$ = 0.17\\\\\hline
\\
& & Lit. & S/N & $\Delta$ [C/Fe] & $\Delta$ [Sr/Fe]  & $\Delta$ [Ba/Fe] & $\Delta$ [Eu/Fe] & $\Delta$ [Ba/Eu]\\
& &      &     & (dex)           &   (dex)           &   (dex)          &   (dex)          &  (dex)  
\\\\\hline
RAVE J010839.6-285701 & 	3 & 	SMSS J010839.58-285701.5 & 	27 & 	$-$0.12 & 	+0.7 & 	$-$0.27 & 	--- & 	--- \\
RAVE J012931.1-160046 & 	1 &	BPS CS 31082-0001, BD-16 251 & 	49 & 	$-$0.09 & 	$-$0.43 & 	+0.16 & 	$-$0.08 & 	+0.24 \\
RAVE J014254.2-503249 & 	1 &	 & 	37 & 	+0.18 & 	$-$0.05 & 	+0.17 & 	+0.02 & 	+0.15 \\
RAVE J014908.0-491143 & 	1 &	HE 0147-4926, CD-49 506 & 	66 & 	+0.02 & 	+0.44 & 	+0.20 & 	+0.26 & 	$-$0.06 \\
RAVE J021110.2-191124 & 	2 &	 & 	39 & 	+0.54 & 	--- & 	--- & 	--- & 	--- \\
RAVE J112651.4-180530 & 	2 &	BD-10 3332 & 	70 & 	+0.25 & 	$-$0.04 & 	+0.39 & 	$-$0.03 & 	+0.41 \\
RAVE J114444.8-112817 & 	2 &	 & 	54 & 	+0.03 & 	+0.04 & 	+0.37 & 	$-$0.23 & 	+0.60 \\
RAVE J133352.8-262354 & 	1 &	CD-33 9514 & 	27 & 	+0.03 & 	+0.01 & 	+0.79 & 	+0.04 & 	+0.75 \\
RAVE J140335.4-333526 & 	1 &	 & 	63 & 	$-$0.34 & 	+0.02 & 	+0.29 & 	$-$0.09 & 	+0.38 \\
RAVE J155829.6-122434 & 	1 &	BD-14 4336 & 	81 & 	$-$0.01 & 	$-$0.33 & 	+0.18 & 	$-$0.21 & 	+0.39 \\
RAVE J160245.0-152102 & 	1 &	 & 	28 & 	+0.13 & 	--- & 	$-$0.21 & 	$-$0.54 & 	+0.33 \\
RAVE J180242.3-440443 & 	1 &	 & 	28 & 	+0.05 & 	+0.84 & 	+0.11 & 	$-$0.05 & 	+0.16 \\
RAVE J183623.2-642812 & 	1 &	 & 	48 & 	$-$0.14 & 	$-$0.23 & 	+0.32 & 	$-$0.13 & 	+0.44 \\
RAVE J192325.2-583341 & 	2 &	 & 	81 & 	$-$0.67 & 	--- & 	+0.80 & 	$-$0.05 & 	+0.67 \\
RAVE J203622.6-071420 & 	1 &	 & 	66 & 	$-$0.10 & 	$-$0.28 & 	+0.36 & 	$-$0.08 & 	+0.44 \\
RAVE J205149.7-615801 & 	1 &	BD-14 5890 , HD 358059 & 	93 & 	+0.02 & 	$-$0.15 & 	+0.53 & 	+0.21 & 	+0.32 \\
RAVE J205609.1-133118 & 	1 &	 & 	58 & 	$-$0.32 & 	$-$0.67 & 	+0.31 & 	$-$0.43 & 	+0.75 \\
RAVE J210642.9-682827 & 	1 &	 & 	21 & 	+0.17 & 	--- & 	+0.14 & 	--- & 	--- \\
RAVE J215118.3-135937 & 	1 &	 & 	66 & 	$-$0.13 & 	$-$0.27 & 	+0.57 & 	$-$0.17 & 	+0.74 \\
RAVE J224927.6-223829 & 	1 &	CD-23 17654 & 	21 & 	$-$0.17 & 	--- & 	+0.09 & 	$-$0.04 & 	+0.13 \\
RAVE J225319.9-224856 & 	1 &	 & 	50 & 	+0.29 & 	$-$0.33 & 	+0.49 & 	--- & 	--- \\
RAVE J230222.9-683323 & 	1 &	 & 	62 & 	+0.12 & 	+0.23 & 	--- & 	--- & 	$-$0.37 \\
 & 	 & 	 & 	 & 	$\mu$ = +0.01 & 	$\mu$ = $-$0.08 & 	$\mu$ = +0.29 & 	$\mu$ = +0.07 & 	$\mu$ = +0.38 \\
 & 	 & 	 & 	 & 	$\sigma$ = 0.23 & 	$\sigma$ = 0.38 & 	$\sigma$ = 0.27 & 	$\sigma$ = 0.17 & 	$\sigma$ = 0.28 
\enddata

\tablecomments{Literature citations are as follows: 1) \citet{Hansen2018} 2) \citet{Sakari2018b} 3)  and \citet{Jacobson2015f}. We compare to the LTE values in \citet{Jacobson2015f}. \mbox{Average} uncertainties for [X/Fe] are as follows: \citet{Hansen2018}: $\sim$0.20-0.40\,dex;  \citet{Sakari2018b}: $\sim$0.05-0.25\,dex;   \citet{Jacobson2015f} $\sim$0.11-0.14\, dex.}

\tablecomments{ $\mu$ refers to biweight location and $\sigma$ refers to biweight scale.}
\end{deluxetable*}

\subsection{Comparison to Prior Stellar-Parameter and Abundance Measurements}

Twenty-two of our sample stars have previous stellar-parameter and elemental-abundance estimates based on for high-resolution spectroscopy reported in the literature.  Table \ref{tab:prev_params} summarizes the differences between the our derived stellar parameters and the abundance estimates from the literature, where available.  

The offset of the residuals in $T_{\rm eff}$ is 54\,K, with a dispersion of 180\,K.

The residuals between surface gravities and microturbulent velocities have small differences, with offsets of 0.10\,dex and 0.06\,km s$^{-1}$, respectively. These parameters exhibit a moderate scatter, with a dispersion of 0.51\,dex and 0.41\,km s$^{-1}$, respectively.

As also shown in Table \ref{tab:prev_params}, residuals in [Fe/H] between our sample and previous studies are relatively small; an offset of +0.13\,dex and a dispersion of 0.17\,dex. 
For [C/Fe] (not including the evolutionary corrections), the difference between our sample stars  and the previous results is quite small. The mean difference is +0.01\,dex, and the dispersion of the residuals is 0.23\,dex.

The n-capture abundance comparisons with previous studies exhibit fair agreement. The offset of the residuals for [Sr/Fe] is $-$0.08\,dex, with a dispersion of 0.38\,dex. The [Ba/Fe] residuals have a larger offset (+0.29\,dex), but a smaller dispersion (0.27\,dex). The [Eu/Fe] measurements agree well with previous studies, with an offset of +0.06\,dex, and a dispersion of 0.17\,dex. The [Ba/Eu] ratio used for determining $s$- and $i$-process enhancement has an offset of +0.38\,dex and a dispersion of 0.28\,dex. 

\section{Summary \& Outlook} \label{sec:sum}

This work presents the results from analyses of high-resolution spectroscopy for 108 MP stars, obtained with the Southern African Large Telescope (SALT). In future work, we plan to obtain better high-resolution spectra for the most interesting (those showing hints of carbon and/or $r$-process-element enhancement) of the $\sim 80$ stars that we were unable to measure stellar parameters and abundances for due to low S/N. 

The stellar parameters we obtained were compared with those from two previous moderate- and low-resolution spectroscopic surveys -- RAVE DR5 \citep{Rave5} and \citet{Placco_2018}, and one based on photometric estimates \citep{Huang_2021}.  The results of the latter two comparisons are improved relative to those from RAVE DR5.

Of the 108 stars we analyzed, 84 are very metal poor (VMP; [Fe/H] $< -2.0$; 53 newly identified), and 3 are extremely metal poor (EMP; [Fe/H] $< -3.0$; 1 newly identified). Elemental abundances were determined for these stars, including carbon abundances. This sample was found to contain 27 stars with evolution-corrected [C/Fe]$_c > +0.50$, and 17 CEMP stars ([C/Fe]$_c$ $> + 0.70$). There are 11 CEMP-$r$ stars (8 newly identified), 1 CEMP-$s$ star (newly identified), 2 possible CEMP-$i$ stars (1 newly identified), and 3 CEMP-no stars (all newly identified) in this work.  We found 11 stars (8 newly identified) that are strongly enhanced in $r$-process elements ($r$-II; [Eu/Fe] $> +0.70$), 38 stars (31 newly identified) that are moderately enhanced in $r$-process elements ($r$-I; $+0.30 < $ [Eu/Fe] $\leq +0.70$), and 1 newly identified limited-$r$ star.  

The newly identified CEMP stars from this sample will be combined with those in the sample of  \citet{Yoon_2016}, and others appearing in the literature since its publication, in order to perform a chemo-dynamical clustering analysis, using methods as described in \citet{Gudin_2021}.  
A similar study for an expanded sample of $r-$process-enhanced stars, including our newly identified examples,  is already underway (Shank et al., in prep.).  We expect these analyses to help refine our understanding of the sites and environments in which these classes of chemically peculiar stars formed in the Milky Way.

\acknowledgments{}

J.Z. K.C.R., and T.C.B acknowledge partial  support  from  grant  PHY  14-30152; Physics Frontier Center/JINA Center for the Evolution of the Elements (JINA-CEE), awarded by the US National  Science  Foundation. KCR is supported by a grant from the Heising-Simons Foundation. The work of V.M.P. is supported by NOIRLab, which is managed by the Association of Universities for Research in Astronomy (AURA) under a cooperative agreement with the National Science Foundation. Y. H. is supported by the Yunnan University grant C176220100006.

All of the observations reported in this paper were obtained with the Southern African Large Telescope (SALT).

\facilities {\textit{Southern African Large Telescope (SALT)} High Resolution Spectrograph (HRS)}
\\

\clearpage
\newpage

\bibliographystyle{aasjournal}
\bibliography{references.bib}

\begin{thebibliography}{}
\expandafter\ifx\csname natexlab\endcsname\relax\def\natexlab#1{#1}\fi
\providecommand{\url}[1]{\href{#1}{#1}}
\providecommand{\dodoi}[1]{doi:~\href{http://doi.org/#1}{\nolinkurl{#1}}}
\providecommand{\doeprint}[1]{\href{http://ascl.net/#1}{\nolinkurl{http://ascl.net/#1}}}
\providecommand{\doarXiv}[1]{\href{https://arxiv.org/abs/#1}{\nolinkurl{https://arxiv.org/abs/#1}}}

\bibitem[{Abbott {et~al.}(2017)Abbott, Abbott, Abbott, Acernese, Ackley, Adams,
  Adams, Addesso, Adhikari, Adya, Affeldt, Afrough, Agarwal, Agathos, Agatsuma,
  Aggarwal, Aguiar, Aiello, Ain, Ajith, Allen, Allen, Allocca, Altin, Amato,
  Ananyeva, Anderson, Anderson, Angelova, Antier, Appert, Arai, Araya, Areeda,
  Arnaud, Arun, Ascenzi, Ashton, Ast, Aston, Astone, Atallah, Aufmuth, Aulbert,
  AultONeal, Austin, Avila-Alvarez, Babak, Bacon, Bader, Bae, Baker,
  Baldaccini, Ballardin, Ballmer, Banagiri, Barayoga, Barclay, Barish, Barker,
  Barkett, Barone, Barr, Barsotti, Barsuglia, Barta, Barthelmy, Bartlett,
  Bartos, Bassiri, Basti, Batch, Bawaj, Bayley, Bazzan, B{\'{e}}csy, Beer,
  Bejger, Belahcene, Bell, Berger, Bergmann, Bero, Berry, Bersanetti,
  Bertolini, Betzwieser, Bhagwat, Bhandare, Bilenko, Billingsley, Billman,
  Birch, Birney, Birnholtz, Biscans, Biscoveanu, Bisht, Bitossi, Biwer,
  Bizouard, Blackburn, Blackman, Blair, Blair, Blair, Bloemen, Bock, Bode,
  Boer, Bogaert, Bohe, Bondu, Bonilla, Bonnand, Boom, Bork, Boschi, Bose,
  Bossie, Bouffanais, Bozzi, Bradaschia, Brady, Branchesi, Brau, Briant,
  Brillet, Brinkmann, Brisson, Brockill, Broida, Brooks, Brown, Brown, Brunett,
  Buchanan, Buikema, Bulik, Bulten, Buonanno, Buskulic, Buy, Byer, Cabero,
  Cadonati, Cagnoli, Cahillane, Bustillo, Callister, Calloni, Camp, Canepa,
  Canizares, Cannon, Cao, Cao, Capano, Capocasa, Carbognani, Caride, Carney,
  Diaz, Casentini, Caudill, Cavagli{\`{a}}, Cavalier, Cavalieri, Cella, Cepeda,
  Cerd{\'{a}}-Dur{\'{a}}n, Cerretani, Cesarini, Chamberlin, Chan, Chao,
  Charlton, Chase, Chassande-Mottin, Chatterjee, Chatziioannou, Cheeseboro,
  Chen, Chen, Chen, Cheng, Chia, Chincarini, Chiummo, Chmiel, Cho, Cho, Chow,
  Christensen, Chu, Chua, Chua, Chung, Chung, Ciani, Ciolfi, Cirelli, Cirone,
  Clara, Clark, Clearwater, Cleva, Cocchieri, Coccia, Cohadon, Cohen, Colla,
  Collette, Cominsky, Jr., Conti, Cooper, Corban, Corbitt,
  Cordero-Carri{\'{o}}n, Corley, Cornish, Corsi, Cortese, Costa, Coughlin,
  Coughlin, Coulon, Countryman, Couvares, Covas, Cowan, Coward, Cowart, Coyne,
  Coyne, Creighton, Creighton, Cripe, Crowder, Cullen, Cumming, Cunningham,
  Cuoco, Canton, D{\'{a}}lya, Danilishin, D'Antonio, Danzmann, Dasgupta, Costa,
  Dattilo, Dave, Davier, Davis, Daw, Day, De, DeBra, Degallaix, Laurentis,
  Del{\'{e}}glise, Pozzo, Demos, Denker, Dent, Pietri, Dergachev, Rosa, DeRosa,
  Rossi, DeSalvo, de~Varona, Devenson, Dhurandhar, D{\'{\i}}az, Fiore,
  Giovanni, Girolamo, Lieto, Pace, Palma, Renzo, Doctor, Dolique, Donovan,
  Dooley, Doravari, Dorrington, Douglas, {\'{A}}lvarez, Downes, Drago,
  Dreissigacker, Driggers, Du, Ducrot, Dupej, Dwyer, Edo, Edwards, Effler,
  Ehrens, Eichholz, Eikenberry, Eisenstein, Essick, Estevez, Etienne, Etzel,
  Evans, Evans, Factourovich, Fafone, Fair, Fairhurst, Fan, Farinon, Farr,
  Farr, Fauchon-Jones, Favata, Fays, Fee, Fehrmann, Feicht, Fejer,
  Fernandez-Galiana, Ferrante, Ferreira, Ferrini, Fidecaro, Finstad, Fiori,
  Fiorucci, Fishbach, Fisher, Fitz-Axen, Flaminio, Fletcher, Fong, Font,
  Forsyth, Forsyth, Fournier, Frasca, Frasconi, Frei, Freise, Frey, Frey,
  Fries, Fritschel, Frolov, Fulda, Fyffe, Gabbard, Gadre, Gaebel, Gair,
  Gammaitoni, Ganija, Gaonkar, Garcia-Quiros, Garufi, Gateley, Gaudio, Gaur,
  Gayathri, Gehrels, Gemme, Genin, Gennai, George, George, Gergely, Germain,
  Ghonge, Ghosh, Ghosh, Ghosh, Giaime, Giardina, Giazotto, Gill, Glover, Goetz,
  Goetz, Gomes, Goncharov, Gonz{\'{a}}lez, Castro, Gopakumar, Gorodetsky,
  Gossan, Gosselin, Gouaty, Grado, Graef, Granata, Grant, Gras, Gray, Greco,
  Green, Gretarsson, Griswold, Groot, Grote, Grunewald, Gruning, Guidi, Guo,
  Gupta, Gupta, Gushwa, Gustafson, Gustafson, Halim, Hall, Hall, Hamilton,
  Hammond, Haney, Hanke, Hanks, Hanna, Hannam, Hannuksela, Hanson, Hardwick,
  Harms, Harry, Harry, Hart, Haster, Haughian, Healy, Heidmann, Heintze,
  Heitmann, Hello, Hemming, Hendry, Heng, Hennig, Heptonstall, Heurs, Hild,
  Hinderer, Hoak, Hofman, Holt, Holz, Hopkins, Horst, Hough, Houston, Howell,
  Hreibi, Hu, Huerta, Huet, Hughey, Husa, Huttner, Huynh-Dinh, Indik, Inta,
  Intini, Isa, Isac, Isi, Iyer, Izumi, Jacqmin, Jani, Jaranowski, Jawahar,
  Jim{\'{e}}nez-Forteza, Johnson, Jones, Jones, Jonker, Ju, Junker, Kalaghatgi,
  Kalogera, Kamai, Kandhasamy, Kang, Kanner, Kapadia, Karki, Karvinen,
  Kasprzack, Katolik, Katsavounidis, Katzman, Kaufer, Kawabe,
  K{\'{e}}f{\'{e}}lian, Keitel, Kemball, Kennedy, Kent, Key, Khalili, Khan,
  Khan, Khan, Khazanov, Kijbunchoo, Kim, Kim, Kim, Kim, Kim, Kim, Kimbrell,
  King, King, Kinley-Hanlon, Kirchhoff, Kissel, Kleybolte, Klimenko, Knowles,
  Koch, Koehlenbeck, Koley, Kondrashov, Kontos, Korobko, Korth, Kowalska,
  Kozak, Krämer, Kringel, Krishnan, Kr{\'{o}}lak, Kuehn, Kumar, Kumar, Kumar,
  Kuo, Kutynia, Kwang, Lackey, Lai, Landry, Lang, Lange, Lantz, Lanza, Larson,
  Lartaux-Vollard, Lasky, Laxen, Lazzarini, Lazzaro, Leaci, Leavey, Lee, Lee,
  Lee, Lee, Lee, Lehmann, Lenon, Leonardi, Leroy, Letendre, Levin, Li, Linker,
  Littenberg, Liu, Lo, Lockerbie, London, Lord, Lorenzini, Loriette, Lormand,
  Losurdo, Lough, Lousto, Lovelace, Lück, Lumaca, Lundgren, Lynch, Ma, Macas,
  Macfoy, Machenschalk, MacInnis, Macleod, Hernandez, Maga{\~{n}}a-Sandoval,
  Zertuche, Magee, Majorana, Maksimovic, Man, Mandic, Mangano, Mansell, Manske,
  Mantovani, Marchesoni, Marion, M{\'{a}}rka, M{\'{a}}rka, Markakis, Markosyan,
  Markowitz, Maros, Marquina, Marsh, Martelli, Martellini, Martin, Martin,
  Martynov, Mason, Massera, Masserot, Massinger, Masso-Reid, Mastrogiovanni,
  Matas, Matichard, Matone, Mavalvala, Mazumder, McCarthy, McClelland,
  McCormick, McCuller, McGuire, McIntyre, McIver, McManus, McNeill, McRae,
  McWilliams, Meacher, Meadors, Mehmet, Meidam, Mejuto-Villa, Melatos, Mendell,
  Mercer, Merilh, Merzougui, Meshkov, Messenger, Messick, Metzdorff, Meyers,
  Miao, Michel, Middleton, Mikhailov, Milano, Miller, Miller, Miller,
  Millhouse, Milovich-Goff, Minazzoli, Minenkov, Ming, Mishra, Mitra,
  Mitrofanov, Mitselmakher, Mittleman, Moffa, Moggi, Mogushi, Mohan, Mohapatra,
  Montani, Moore, Moraru, Moreno, Morriss, Mours, Mow-Lowry, Mueller, Muir,
  Mukherjee, Mukherjee, Mukherjee, Mukund, Mullavey, Munch, Mu{\~{n}}iz,
  Muratore, Murray, Napier, Nardecchia, Naticchioni, Nayak, Neilson, Nelemans,
  Nelson, Nery, Neunzert, Nevin, Newport, Newton, Ng, Nguyen, Nguyen, Nichols,
  Nielsen, Nissanke, Nitz, Noack, Nocera, Nolting, North, Nuttall, Oberling,
  O'Dea, Ogin, Oh, Oh, Ohme, Okada, Oliver, Oppermann, Oram, O'Reilly,
  Ormiston, Ortega, O'Shaughnessy, Ossokine, Ottaway, Overmier, Owen, Pace,
  Page, Page, Pai, Pai, Palamos, Palashov, Palomba, Pal-Singh, Pan, Pan, Pang,
  Pang, Pankow, Pannarale, Pant, Paoletti, Paoli, Papa, Parida, Parker,
  Pascucci, Pasqualetti, Passaquieti, Passuello, Patil, Patricelli, Pearlstone,
  Pedraza, Pedurand, Pekowsky, Pele, Penn, Perez, Perreca, Perri, Pfeiffer,
  Phelps, Piccinni, Pichot, Piergiovanni, Pierro, Pillant, Pinard, Pinto,
  Pirello, Pitkin, Poe, Poggiani, Popolizio, Porter, Post, Powell, Prasad,
  Pratt, Pratten, Predoi, Prestegard, Price, Prijatelj, Principe, Privitera,
  Prodi, Prokhorov, Puncken, Punturo, Puppo, Pürrer, Qi, Quetschke, Quintero,
  Quitzow-James, Raab, Rabeling, Radkins, Raffai, Raja, Rajan, Rajbhandari,
  Rakhmanov, Ramirez, Ramos-Buades, Rapagnani, Raymond, Razzano, Read,
  Regimbau, Rei, Reid, Reitze, Ren, Reyes, Ricci, Ricker, Rieger, Riles, Rizzo,
  Robertson, Robie, Robinet, Rocchi, Rolland, Rollins, Roma, Romano, Romel,
  Romie, Rosi{\'{n}}ska, Ross, Rowan, Rüdiger, Ruggi, Rutins, Ryan, Sachdev,
  Sadecki, Sadeghian, Sakellariadou, Salconi, Saleem, Salemi, Samajdar, Sammut,
  Sampson, Sanchez, Sanchez, Sanchis-Gual, Sandberg, Sanders, Sassolas,
  Sathyaprakash, Saulson, Sauter, Savage, Sawadsky, Schale, Scheel, Scheuer,
  Schmidt, Schmidt, Schnabel, Schofield, Schönbeck, Schreiber, Schuette,
  Schulte, Schutz, Schwalbe, Scott, Scott, Seidel, Sellers, Sengupta, Sentenac,
  Sequino, Sergeev, Shaddock, Shaffer, Shah, Shahriar, Shaner, Shao, Shapiro,
  Shawhan, Sheperd, Shoemaker, Shoemaker, Siellez, Siemens, Sieniawska, Sigg,
  Silva, Singer, Singh, Singhal, Sintes, Slagmolen, Smith, Smith, Smith,
  Somala, Son, Sonnenberg, Sorazu, Sorrentino, Souradeep, Spencer, Srivastava,
  Staats, Staley, Steinke, Steinlechner, Steinlechner, Steinmeyer, Stevenson,
  Stone, Stops, Strain, Stratta, Strigin, Strunk, Sturani, Stuver,
  Summerscales, Sun, Sunil, Suresh, Sutton, Swinkels, Szczepa{\'{n}}czyk,
  Tacca, Tait, Talbot, Talukder, Tanner, T{\'{a}}pai, Taracchini, Tasson,
  Taylor, Taylor, Tewari, Theeg, Thies, Thomas, Thomas, Thomas, Thorne, Thorne,
  Thrane, Tiwari, Tiwari, Tokmakov, Toland, Tonelli, Tornasi,
  Torres-Forn{\'{e}}, Torrie, Töyrä, Travasso, Traylor, Trinastic, Tringali,
  Trozzo, Tsang, Tse, Tso, Tsukada, Tsuna, Tuyenbayev, Ueno, Ugolini,
  Unnikrishnan, Urban, Usman, Vahlbruch, Vajente, Valdes, van Bakel, van
  Beuzekom, van~den Brand, Broeck, Vander-Hyde, van~der Schaaf, van Heijningen,
  van Veggel, Vardaro, Varma, Vass, Vas{\'{u}}th, Vecchio, Vedovato, Veitch,
  Veitch, Venkateswara, Venugopalan, Verkindt, Vetrano, Vicer{\'{e}}, Viets,
  Vinciguerra, Vine, Vinet, Vitale, Vo, Vocca, Vorvick, Vyatchanin, Wade, Wade,
  Wade, Walet, Walker, Wallace, Walsh, Wang, Wang, Wang, Wang, Wang, Ward,
  Warner, Was, Watchi, Weaver, Wei, Weinert, Weinstein, Weiss, Wen, Wessel,
  Wessels, Westerweck, Westphal, Wette, Whelan, Whitcomb, Whiting, Whittle,
  Wilken, Williams, Williams, Williamson, Willis, Willke, Wimmer, Winkler,
  Wipf, Wittel, Woan, Woehler, Wofford, Wong, Worden, Wright, Wu, Wysocki,
  Xiao, Yamamoto, Yancey, Yang, Yap, Yazback, Yu, Yu, Yvert, Zadro{\.{z}}ny,
  Zanolin, Zelenova, Zendri, Zevin, Zhang, Zhang, Zhang, Zhang, Zhao, Zhou,
  Zhou, Zhu, Zhu, Zimmerman, Zucker, Zweizig, Wilson-Hodge, Bissaldi,
  Blackburn, Briggs, Burns, Cleveland, Connaughton, Gibby, Giles, Goldstein,
  Hamburg, Jenke, Hui, Kippen, Kocevski, McBreen, Meegan, Paciesas, Poolakkil,
  Preece, Racusin, Roberts, Stanbro, Veres, von Kienlin, Savchenko, Ferrigno,
  Kuulkers, Bazzano, Bozzo, Brandt, Chenevez, Courvoisier, Diehl, Domingo,
  Hanlon, Jourdain, Laurent, Lebrun, Lutovinov, Martin-Carrillo, Mereghetti,
  Natalucci, Rodi, Roques, Sunyaev, Ubertini, Aartsen, Ackermann, Adams,
  Aguilar, Ahlers, Ahrens, Samarai, Altmann, Andeen, Anderson, Ansseau, Anton,
  Argüelles, Auffenberg, Axani, Bagherpour, Bai, Barron, Barwick, Baum, Bay,
  Beatty, Tjus, Bernardini, Besson, Binder, Bindig, Blaufuss, Blot, Bohm,
  Börner, Bos, Bose, Böser, Botner, Bourbeau, Bourbeau, Bradascio, Braun,
  Brayeur, Brenzke, Bretz, Bron, Brostean-Kaiser, Burgman, Carver, Casey,
  Casier, Cheung, Chirkin, Christov, Clark, Classen, Coenders, Collin, Conrad,
  Cowen, Cross, Day, de~Andr{\'{e}}, Clercq, DeLaunay, Dembinski, Ridder,
  Desiati, de~Vries, de~Wasseige, de~With, DeYoung, D{\'{\i}}az-V{\'{e}}lez,
  di~Lorenzo, Dujmovic, Dumm, Dunkman, Dvorak, Eberhardt, Ehrhardt, Eichmann,
  Eller, Evenson, Fahey, Fazely, Felde, Filimonov, Finley, Flis, Franckowiak,
  Friedman, Fuchs, Gaisser, Gallagher, Gerhardt, Ghorbani, Giang, Glauch,
  Glüsenkamp, Goldschmidt, Gonzalez, Grant, Griffith, Haack, Hallgren, Halzen,
  Hanson, Hebecker, Heereman, Helbing, Hellauer, Hickford, Hignight, Hill,
  Hoffman, Hoffmann, Hokanson-Fasig, Hoshina, Huang, Huber, Hultqvist,
  Hünnefeld, In, Ishihara, Jacobi, Japaridze, Jeong, Jero, Jones, Kalaczynski,
  Kang, Kappes, Karg, Karle, Kauer, Keivani, Kelley, Kheirandish, Kim, Kim,
  Kintscher, Kiryluk, Kittler, Klein, Kohnen, Koirala, Kolanoski, Köpke,
  Kopper, Kopper, Koschinsky, Koskinen, Kowalski, Krings, Kroll, Krückl,
  Kunnen, Kunwar, Kurahashi, Kuwabara, Kyriacou, Labare, Lanfranchi, Larson,
  Lauber, Lesiak-Bzdak, Leuermann, Liu, Lu, Lünemann, Luszczak, Madsen, Maggi,
  Mahn, Mancina, Maruyama, Mase, Maunu, McNally, Meagher, Medici, Meier, Menne,
  Merino, Meures, Miarecki, Micallef, Moment{\'{e}}, Montaruli, Moore, Moulai,
  Nahnhauer, Nakarmi, Naumann, Neer, Niederhausen, Nowicki, Nygren, Pollmann,
  Olivas, O'Murchadha, Palczewski, Pandya, Pankova, Peiffer, Pepper, de~los
  Heros, Pieloth, Pinat, Price, Przybylski, Raab, Rädel, Rameez, Rawlins, Rea,
  Reimann, Relethford, Relich, Resconi, Rhode, Richman, Robertson, Rongen,
  Rott, Ruhe, Ryckbosch, Rysewyk, Sälzer, Herrera, Sandrock, Sandroos,
  Santander, Sarkar, Sarkar, Satalecka, Schlunder, Schmidt, Schneider,
  Schoenen, Schöneberg, Schumacher, Seckel, Seunarine, Soedingrekso, Soldin,
  Song, Spiczak, Spiering, Stachurska, Stamatikos, Stanev, Stasik, Stettner,
  Steuer, Stezelberger, Stokstad, Stössl, Strotjohann, Stuttard, Sullivan,
  Sutherland, Taboada, Tatar, Tenholt, Ter-Antonyan, Terliuk,
  Te{\v{s}}i{\'{c}}, Tilav, Toale, Tobin, Toscano, Tosi, Tselengidou, Tung,
  Turcati, Turley, Ty, Unger, Usner, Vandenbroucke, Driessche, van Eijndhoven,
  Vanheule, van Santen, Vehring, Vogel, Vraeghe, Walck, Wallace, Wallraff,
  Wandler, Wandkowsky, Waza, Weaver, Weiss, Wendt, Werthebach, Whelan, Wiebe,
  Wiebusch, Wille, Williams, Wills, Wolf, Wood, Woolsey, Woschnagg, Xu, Xu, Xu,
  Yanez, Yodh, Yoshida, Yuan, Zoll, Balasubramanian, Mate, Bhalerao,
  Bhattacharya, Vibhute, Dewangan, Rao, Vadawale, Svinkin, Hurley, Aptekar,
  Frederiks, Golenetskii, Kozlova, Lysenko, Oleynik, Tsvetkova, Ulanov, Cline,
  Li, Xiong, Zhang, Lu, Song, Cao, Chang, Chen, Chen, Chen, Chen, Chen, Chen,
  Cui, Cui, Deng, Dong, Du, Fu, Gao, Gao, Gao, Ge, Gu, Guan, Guo, Han, Hu,
  Huang, Huo, Jia, Jiang, Jiang, Jin, Jin, Li, Li, Li, Li, Li, Li, Li, Li, Li,
  Li, Li, Liang, Liao, Liu, Liu, Liu, Liu, Liu, Liu, Liu, Lu, Lu, Luo, Ma,
  Meng, Nang, Nie, Ou, Qu, Sai, Sun, Tan, Tao, Tao, Tuo, Wang, Wang, Wang,
  Wang, Wang, Wen, Wu, Wu, Xiao, Xu, Xu, Yan, Yang, Yang, Yang, Zhang, Zhang,
  Zhang, Zhang, Zhang, Zhang, Zhang, Zhang, Zhang, Zhang, Zhang, Zhang, Zhang,
  Zhang, Zhang, Zhang, Zhang, Zhang, Zhao, Zhao, Zhao, Zheng, Zhu, Zhu, Zou,
  Albert, Andr{\'{e}}, Anghinolfi, Ardid, Aubert, Aublin, Avgitas, Baret,
  Barrios-Mart{\'{\i}}, Basa, Belhorma, Bertin, Biagi, Bormuth, Bourret,
  Bouwhuis, Br{\^{a}}nza{\c{s}}, Bruijn, Brunner, Busto, Capone, Caramete,
  Carr, Celli, Moursli, Chiarusi, Circella, Coelho, Coleiro, Coniglione,
  Costantini, Coyle, Creusot, D{\'{\i}}az, Deschamps, Bonis, Distefano, Palma,
  Domi, Donzaud, Dornic, Drouhin, Eberl, Bojaddaini, Khayati, Elsässer,
  Enzenhöfer, Ettahiri, Fassi, Felis, Fusco, Gay, Giordano, Glotin,
  Gr{\'{e}}goire, Ruiz, Graf, Hallmann, van Haren, Heijboer, Hello,
  Hern{\'{a}}ndez-Rey, Hössl, Hofestädt, Hugon, Illuminati, James, de~Jong,
  Jongen, Kadler, Kalekin, Katz, Kiessling, Kouchner, Kreter, Kreykenbohm,
  Kulikovskiy, Lachaud, Lahmann, Lef{\`{e}}vre, Leonora, Lotze, Loucatos,
  Marcelin, Margiotta, Marinelli, Mart{\'{\i}}nez-Mora, Mele, Melis, Michael,
  Migliozzi, Moussa, Navas, Nezri, Organokov, P{\u{a}}v{\u{a}}la{\c{s}},
  Pellegrino, Perrina, Piattelli, Popa, Pradier, Quinn, Racca, Riccobene,
  S{\'{a}}nchez-Losa, Salda{\~{n}}a, Salvadori, Samtleben, Sanguineti,
  Sapienza, Sieger, Spurio, Stolarczyk, Taiuti, Tayalati, Trovato, Turpin,
  Tönnis, Vallage, Elewyck, Versari, Vivolo, Vizzoca, Wilms, Zornoza,
  Z{\'{u}}{\~{n}}iga, Beardmore, Breeveld, Burrows, Cenko, Cusumano,
  D'A{\`{\i}}, de~Pasquale, Emery, Evans, Giommi, Gronwall, Kennea, Krimm,
  Kuin, Lien, Marshall, Melandri, Nousek, Oates, Osborne, Pagani, Page, Palmer,
  Perri, Siegel, Sbarufatti, Tagliaferri, Tohuvavohu, Tavani, Verrecchia,
  Bulgarelli, Evangelista, Pacciani, Feroci, Pittori, Giuliani, Monte,
  Donnarumma, Argan, Trois, Ursi, Cardillo, Piano, Longo, Lucarelli,
  Munar-Adrover, Fuschino, Labanti, Marisaldi, Minervini, Fioretti,
  Parmiggiani, Gianotti, Trifoglio, Persio, Antonelli, Barbiellini, Caraveo,
  Cattaneo, Costa, Colafrancesco, D'Amico, Ferrari, Morselli, Paoletti,
  Picozza, Pilia, Rappoldi, Soffitta, Vercellone, Foley, Coulter, Kilpatrick,
  Drout, Piro, Shappee, Siebert, Simon, Ulloa, Kasen, Madore, Murguia-Berthier,
  Pan, Prochaska, Ramirez-Ruiz, Rest, Rojas-Bravo, Berger, Soares-Santos,
  Annis, Alexander, Allam, Balbinot, Blanchard, Brout, Butler, Chornock, Cook,
  Cowperthwaite, Diehl, Drlica-Wagner, Drout, Durret, Eftekhari, Finley, Fong,
  Frieman, Fryer, Garc{\'{\i}}a-Bellido, Gruendl, Hartley, Herner, Kessler,
  Lin, Lopes, Louren{\c{c}}o, Margutti, Marshall, Matheson, Medina, Metzger,
  Mu{\~{n}}oz, Muir, Nicholl, Nugent, Palmese, Paz-Chinch{\'{o}}n, Quataert,
  Sako, Sauseda, Schlegel, Scolnic, Secco, Smith, Sobreira, Villar, Vivas,
  Wester, Williams, Yanny, Zenteno, Zhang, Abbott, Banerji, Bechtol,
  Benoit-L{\'{e}}vy, Bertin, Brooks, Buckley-Geer, Burke, Capozzi, Rosell,
  Kind, Castander, Crocce, Cunha, D'Andrea, da~Costa, Davis, DePoy, Desai,
  Dietrich, Eifler, Fernandez, Flaugher, Fosalba, Gaztanaga, Gerdes,
  Giannantonio, Goldstein, Gruen, Gschwend, Gutierrez, Honscheid, James,
  Jeltema, Johnson, Johnson, Kent, Krause, Kron, Kuehn, Lahav, Lima, Maia,
  March, Martini, McMahon, Menanteau, Miller, Miquel, Mohr, Nichol, Ogando,
  Plazas, Romer, Roodman, Rykoff, Sanchez, Scarpine, Schindler, Schubnell,
  Sevilla-Noarbe, Sheldon, Smith, Smith, Stebbins, Suchyta, Swanson, Tarle,
  Thomas, Troxel, Tucker, Vikram, Walker, Wechsler, Weller, Carlin, Gill, Li,
  Marriner, Neilsen, Haislip, Kouprianov, Reichart, Sand, Tartaglia, Valenti,
  Yang, Benetti, Brocato, Campana, Cappellaro, Covino, D'Avanzo, D'Elia,
  Getman, Ghirlanda, Ghisellini, Limatola, Nicastro, Palazzi, Pian,
  Piranomonte, Possenti, Rossi, Salafia, Tomasella, Amati, Antonelli,
  Bernardini, Bufano, Capaccioli, Casella, Dadina, Cesare, Paola, Giuffrida,
  Giunta, Israel, Lisi, Maiorano, Mapelli, Masetti, Pescalli, Pulone,
  Salvaterra, Schipani, Spera, Stamerra, Stella, Testa, Turatto, Vergani,
  Aresu, Bachetti, Buffa, Burgay, Buttu, Caria, Carretti, Casasola, Castangia,
  Carboni, Casu, Concu, Corongiu, Deiana, Egron, Fara, Gaudiomonte, Gusai,
  Ladu, Loru, Leurini, Marongiu, Melis, Melis, Migoni, Milia, Navarrini,
  Orlati, Ortu, Palmas, Pellizzoni, Perrodin, Pisanu, Poppi, Righini, Saba,
  Serra, Serrau, Stagni, Surcis, Vacca, Vargiu, Hunt, Jin, Klose, Kouveliotou,
  Mazzali, M{\o}ller, Nava, Piran, Selsing, Vergani, Wiersema, Toma, Higgins,
  Mundell, di~Serego~Alighieri, G{\'{o}}tz, Gao, Gomboc, Kaper, Kobayashi,
  Kopac, Mao, Starling, Steele, van~der Horst, Acero, Atwood, Baldini,
  Barbiellini, Bastieri, Berenji, Bellazzini, Bissaldi, Blandford, Bloom,
  Bonino, Bottacini, Bregeon, Buehler, Buson, Cameron, Caputo, Caraveo,
  Cavazzuti, Chekhtman, Cheung, Chiang, Ciprini, Cohen-Tanugi, Cominsky,
  Costantin, Cuoco, D'Ammando, de~Palma, Digel, Lalla, Mauro, Venere, Dubois,
  Fegan, Focke, Franckowiak, Fukazawa, Funk, Fusco, Gargano, Gasparrini,
  Giglietto, Giordano, Giroletti, Glanzman, Green, Grondin, Guillemot, Guiriec,
  Harding, Horan, J{\'{o}}hannesson, Kamae, Kensei, Kuss, Mura, Latronico,
  Lemoine-Goumard, Longo, Loparco, Lovellette, Lubrano, Magill, Maldera,
  Manfreda, Mazziotta, McEnery, Meyer, Michelson, Mirabal, Monzani, Moretti,
  Morselli, Moskalenko, Negro, Nuss, Ojha, Omodei, Orienti, Orlando,
  Palatiello, Paliya, Paneque, Pesce-Rollins, Piron, Porter, Principe,
  Rain{\`{o}}, Rando, Razzano, Razzaque, Reimer, Reimer, Reposeur, Rochester,
  Parkinson, Sgr{\`{o}}, Siskind, Spada, Spandre, Suson, Takahashi, Tanaka,
  Thayer, Thayer, Thompson, Tibaldo, Torres, Torresi, Troja, Venters, Vianello,
  Zaharijas, Allison, Bannister, Dobie, Kaplan, Lenc, Lynch, Murphy, Sadler,
  Hotan, James, Oslowski, Raja, Shannon, Whiting, Arcavi, Howell, McCully,
  Hosseinzadeh, Hiramatsu, Poznanski, Barnes, Zaltzman, Vasylyev, Maoz, Cooke,
  Bailes, Wolf, Deller, Lidman, Wang, Gendre, Andreoni, Ackley, Pritchard,
  Bessell, Chang, Möller, Onken, Scalzo, Ridden-Harper, Sharp, Tucker,
  Farrell, Elmer, Johnston, Krishnan, Keane, Green, Jameson, Hu, Ma, Sun, Wu,
  Wang, Shang, Hu, Ashley, Yuan, Li, Tao, Zhu, Zhang, Suntzeff, Zhou, Yang,
  Orange, Morris, Cucchiara, Giblin, Klotz, Staff, Thierry, Schmidt, Tanvir,
  Levan, Cano, de~Ugarte-Postigo, Gonz{\'{a}}lez-Fern{\'{a}}ndez, Greiner,
  Hjorth, Irwin, Krühler, Mandel, Milvang-Jensen, O'Brien, Rol, Rosetti,
  Rosswog, Rowlinson, Steeghs, Thöne, Ulaczyk, Watson, Bruun, Cutter, Jaimes,
  Fujii, Fruchter, Gompertz, Jakobsson, Hodosan, J{\`{e}}rgensen, Kangas, Kann,
  Rabus, Schr{\o}der, Stanway, Wijers, Lipunov, Gorbovskoy, Kornilov, Tyurina,
  Balanutsa, Kuznetsov, Vlasenko, Podesta, Lopez, Podesta, Levato, Saffe,
  Mallamaci, Budnev, Gress, Kuvshinov, Gorbunov, Vladimirov, Zimnukhov,
  Gabovich, Yurkov, Sergienko, Rebolo, Serra-Ricart, Tlatov, Ishmuhametova,
  Abe, Aoki, Aoki, Asakura, Baar, Barway, Bond, Doi, Finet, Fujiyoshi,
  Furusawa, Honda, Itoh, Kanda, Kawabata, Kawabata, Kim, Koshida, Kuroda, Lee,
  Liu, Matsubayashi, Miyazaki, Morihana, Morokuma, Motohara, Murata, Nagai,
  Nagashima, Nagayama, Nakaoka, Nakata, Ohsawa, Ohshima, Ohta, Okita, Saito,
  Saito, Sako, Sekiguchi, Sumi, Tajitsu, Takahashi, Takayama, Tamura, Tanaka,
  Tanaka, Terai, Tominaga, Tristram, Uemura, Utsumi, Yamaguchi, Yasuda,
  Yoshida, Zenko, Adams, Anupama, Bally, Barway, Bellm, Blagorodnova, Cannella,
  Chandra, Chatterjee, Clarke, Cobb, Cook, Copperwheat, De, Emery, Feindt,
  Foster, Fox, Frail, Fremling, Frohmaier, Garcia, Ghosh, Giacintucci, Goobar,
  Gottlieb, Grefenstette, Hallinan, Harrison, Heida, Helou, Ho, Horesh,
  Hotokezaka, Ip, Itoh, Jacobs, Jencson, Kasen, Kasliwal, Kassim, Kim, Kiran,
  Kuin, Kulkarni, Kupfer, Lau, Madsen, Mazzali, Miller, Miyasaka, Mooley,
  Myers, Nakar, Ngeow, Nugent, Ofek, Palliyaguru, Pavana, Perley, Peters, Pike,
  Piran, Qi, Quimby, Rana, Rosswog, Rusu, Sadler, Sistine, Sollerman, Xu, Yan,
  Yatsu, Yu, Zhang, Zhao, Chambers, Huber, Schultz, Bulger, Flewelling,
  Magnier, Lowe, Wainscoat, Waters, Willman, Ebisawa, Hanyu, Harita, Hashimoto,
  Hidaka, Hori, Ishikawa, Isobe, Iwakiri, Kawai, Kawai, Kawamuro, Kawase,
  Kitaoka, Makishima, Matsuoka, Mihara, Morita, Morita, Nakahira, Nakajima,
  Nakamura, Negoro, Oda, Sakamaki, Sasaki, Serino, Shidatsu, Shimomukai,
  Sugawara, Sugita, Sugizaki, Tachibana, Takao, Tanimoto, Tomida, Tsuboi,
  Tsunemi, Ueda, Ueno, Yamada, Yamaoka, Yamauchi, Yatabe, Yoneyama, Yoshii,
  Coward, Crisp, Macpherson, Andreoni, Laugier, Noysena, Klotz, Gendre,
  Thierry, Turpin, Im, Choi, Kim, Yoon, Lim, Lee, Lee, Kim, Ko, Joe, Kwon, Kim,
  Lim, Choi, Fynbo, Malesani, Xu, Smartt, Jerkstrand, Kankare, Sim, Fraser,
  Inserra, Maguire, Leloudas, Magee, Shingles, Smith, Young, Kotak, Gal-Yam,
  Lyman, Homan, Agliozzo, Anderson, Angus, Ashall, Barbarino, Bauer, Berton,
  Botticella, Bulla, Cannizzaro, Cartier, Cikota, Clark, Cia, Valle, Dennefeld,
  Dessart, Dimitriadis, Elias-Rosa, Firth, Flörs, Frohmaier, Galbany,
  Gonz{\'{a}}lez-Gait{\'{a}}n, Gromadzki, Guti{\'{e}}rrez, Hamanowicz,
  Harmanen, Heintz, Hernandez, Hodgkin, Hook, Izzo, James, Jonker, Kerzendorf,
  Kostrzewa-Rutkowska, Kromer, Kuncarayakti, Lawrence, Manulis, Mattila,
  McBrien, Müller, Nordin, O'Neill, Onori, Palmerio, Pastorello, Patat,
  Pignata, Podsiadlowski, Razza, Reynolds, Roy, Ruiter, Rybicki, Salmon, Pumo,
  Prentice, Seitenzahl, Smith, Sollerman, Sullivan, Szegedi, Taddia,
  Taubenberger, Terreran, Soelen, Vos, Walton, Wright, Wyrzykowski, Yaron,
  Chen, Krühler, Schady, Wiseman, Greiner, Rau, Schweyer, Klose, Guelbenzu,
  Palliyaguru, Shara, Williams, Vaisanen, Potter, Colmenero, Crawford, Buckley,
  Mao, D{\'{\i}}az, Macri, Lambas, de~Oliveira, Castell{\'{o}}n, Ribeiro,
  S{\'{a}}nchez, Schoenell, Abramo, Akras, Alcaniz, Artola, Beroiz, Bonoli,
  Cabral, Camuccio, Chavushyan, Coelho, Colazo, Costa-Duarte, Larenas, Romero,
  Dultzin, Fern{\'{a}}ndez, Garc{\'{\i}}a, Girardini, Gon{\c{c}}alves,
  Gon{\c{c}}alves, Gurovich, Jim{\'{e}}nez-Teja, Kanaan, Lares, de~Oliveira,
  L{\'{o}}pez-Cruz, Melia, Molino, Padilla, Pe{\~{n}}uela, Placco,
  Qui{\~{n}}ones, Rivera, Renzi, Riguccini, R{\'{\i}}os-L{\'{o}}pez, Rodriguez,
  Sampedro, Schneiter, Sodr{\'{e}}, Starck, Torres-Flores, Tornatore,
  Zadro{\.{z}}ny, Castillo, Castro-Tirado, Tello, Hu, Zhang, Cunniffe,
  Castell{\'{o}}n, Hiriart, Caballero-Garc{\'{\i}}a, Jel{\'{\i}}nek,
  Kub{\'{a}}nek, del Pulgar, Park, Jeong, Cer{\'{o}}n, Pandey, Yock, Querel,
  Fan, Wang, Beardsley, Brown, Crosse, Emrich, Franzen, Gaensler, Horsley,
  Johnston-Hollitt, Kenney, Morales, Pallot, Sokolowski, Steele, Tingay, Trott,
  Walker, Wayth, Williams, Wu, Yoshida, Sakamoto, Kawakubo, Yamaoka, Takahashi,
  Asaoka, Ozawa, Torii, Shimizu, Tamura, Ishizaki, Cherry, Ricciarini,
  Penacchioni, Marrocchesi, Pozanenko, Volnova, Mazaeva, Minaev, Krugov,
  Kusakin, Reva, Moskvitin, Rumyantsev, Inasaridze, Klunko, Tungalag, Schmalz,
  Burhonov, Abdalla, Abramowski, Aharonian, Benkhali, Angüner, Arakawa,
  Arrieta, Aubert, Backes, Balzer, Barnard, Becherini, Tjus, Berge, Bernhard,
  Bernlöhr, Blackwell, Böttcher, Boisson, Bolmont, Bonnefoy, Bordas, Bregeon,
  Brun, Brun, Bryan, Büchele, Bulik, Capasso, Caroff, Carosi, Casanova,
  Cerruti, Chakraborty, Chaves, Chen, Chevalier, Colafrancesco, Condon, Conrad,
  Davids, Decock, Deil, Devin, deWilt, Dirson, Djannati-Ataï, Donath, Drury,
  Dutson, Dyks, Edwards, Egberts, Emery, Ernenwein, Eschbach, Farnier, Fegan,
  Fernandes, Fiasson, Fontaine, Funk, Füssling, Gabici, Gallant, Garrigoux,
  Gat{\'{e}}, Giavitto, Giebels, Glawion, Glicenstein, Gottschall, Grondin,
  Hahn, Haupt, Hawkes, Heinzelmann, Henri, Hermann, Hinton, Hofmann, Hoischen,
  Holch, Holler, Horns, Ivascenko, Iwasaki, Jacholkowska, Jamrozy, Jankowsky,
  Jankowsky, Jingo, Jouvin, Jung-Richardt, Kastendieck, Katarzy{\'{n}}ski,
  Katsuragawa, Kerszberg, Khangulyan, Kh{\'{e}}lifi, King, Klepser, Klochkov,
  Klu{\'{z}}niak, Komin, Kosack, Krakau, Kraus, Krüger, Laffon, Lamanna, Lau,
  Lees, Lefaucheur, Lemi{\`{e}}re, Lemoine-Goumard, Lenain, Leser, Lohse,
  Lorentz, Liu, Lypova, Malyshev, Marandon, Marcowith, Mariaud, Marx, Maurin,
  Maxted, Mayer, Meintjes, Meyer, Mitchell, Moderski, Mohamed, Mohrmann,
  Mor{\aa}, Moulin, Murach, Nakashima, de~Naurois, Ndiyavala, Niederwanger,
  Niemiec, Oakes, O'Brien, Odaka, Ohm, Ostrowski, Oya, Padovani, Panter,
  Parsons, Pekeur, Pelletier, Perennes, Petrucci, Peyaud, Piel, Pita, Poireau,
  Poon, Prokhorov, Prokoph, Pühlhofer, Punch, Quirrenbach, Raab, Rauth,
  Reimer, Reimer, Renaud, de~los Reyes, Rieger, Rinchiuso, Romoli, Rowell,
  Rudak, Rulten, Sahakian, Saito, Sanchez, Santangelo, Sasaki, Schlickeiser,
  Schüssler, Schulz, Schwanke, Schwemmer, Seglar-Arroyo, Settimo, Seyffert,
  Shafi, Shilon, Shiningayamwe, Simoni, Sol, Spanier, Spir-Jacob, Stawarz,
  Steenkamp, Stegmann, Steppa, Sushch, Takahashi, Tavernet, Tavernier, Taylor,
  Terrier, Tibaldo, Tiziani, Tluczykont, Trichard, Tsirou, Tsuji, Tuffs,
  Uchiyama, van~der Walt, van Eldik, van Rensburg, van Soelen, Vasileiadis,
  Veh, Venter, Viana, Vincent, Vink, Voisin, Völk, Vuillaume, Wadiasingh,
  Wagner, Wagner, Wagner, White, Wierzcholska, Willmann, Wörnlein, Wouters,
  Yang, Zaborov, Zacharias, Zanin, Zdziarski, Zech, Zefi, Ziegler, Zorn,
  {\.{Z}}ywucka, Fender, Broderick, Rowlinson, Wijers, Stewart, ter Veen,
  Shulevski, Kavic, Simonetti, League, Tsai, Obenberger, Nathaniel, Taylor,
  Dowell, Liebling, Estes, Lippert, Sharma, Vincent, Farella, Abeysekara,
  Albert, Alfaro, Alvarez, Arceo, Arteaga-Vel{\'{a}}zquez, Rojas, Solares,
  Barber, Gonzalez, Becerril, Belmont-Moreno, BenZvi, Berley, Bernal, Braun,
  Brisbois, Caballero-Mora, Capistr{\'{a}}n, Carrami{\~{n}}ana, Casanova,
  Castillo, Cotti, Cotzomi, de~Le{\'{o}}n, Le{\'{o}}n, la~Fuente, Hernandez,
  Dichiara, Dingus, DuVernois, D{\'{\i}}az-V{\'{e}}lez, Ellsworth, Engel,
  Enr{\'{\i}}quez-Rivera, Fiorino, Fleischhack, Fraija,
  Garc{\'{\i}}a-Gonz{\'{a}}lez, Garfias, Gerhardt, Mu{\~{n}}oz, Gonz{\'{a}}lez,
  Goodman, Hampel-Arias, Harding, Hernandez, Hernandez-Almada, Hona,
  Hüntemeyer, Iriarte, Jardin-Blicq, Joshi, Kaufmann, Kieda, Lara, Lauer,
  Lennarz, Vargas, Linnemann, Longinotti, Raya, Luna-Garc{\'{\i}}a,
  L{\'{o}}pez-Coto, Malone, Marinelli, Martinez, Martinez-Castellanos,
  Mart{\'{\i}}nez-Castro, Mart{\'{\i}}nez-Huerta, Matthews, Miranda-Romagnoli,
  Moreno, Mostaf{\'{a}}, Nellen, Newbold, Nisa, Noriega-Papaqui, Pelayo, Pretz,
  P{\'{e}}rez-P{\'{e}}rez, Ren, Rho, Rivi{\`{e}}re, Rosa-Gonz{\'{a}}lez,
  Rosenberg, Ruiz-Velasco, Salazar, Greus, Sandoval, Schneider, Schoorlemmer,
  Sinnis, Smith, Springer, Surajbali, Tibolla, Tollefson, Torres, Ukwatta,
  Weisgarber, Westerhoff, Wisher, Wood, Yapici, Yodh, Younk, Zhou,
  {\'{A}}lvarez, Aab, Abreu, Aglietta, Albuquerque, Albury, Allekotte, Almela,
  Castillo, Alvarez-Mu{\~{n}}iz, Anastasi, Anchordoqui, Andrada, Andringa,
  Aramo, Arsene, Asorey, Assis, Avila, Badescu, Balaceanu, Barbato, Luz,
  Becker, Bellido, Berat, Bertaina, Bertou, Biermann, Biteau, Blaess, Blanco,
  Blazek, Bleve, Boh{\'{a}}{\v{c}}ov{\'{a}}, Bonifazi, Borodai, Botti, Brack,
  Brancus, Bretz, Bridgeman, Briechle, Buchholz, Bueno, Buitink, Buscemi,
  Caballero-Mora, Caccianiga, Cancio, Canfora, Caruso, Castellina, Catalani,
  Cataldi, Cazon, Chavez, Chinellato, Chudoba, Clay, Cerutti, Colalillo,
  Coleman, Collica, Coluccia, Concei{\c{c}}{\~{a}}o, Consolati, Contreras,
  Cooper, Coutu, Covault, Cronin, D'Amico, Daniel, Dasso, Daumiller, Dawson,
  Day, de~Almeida, de~Jong, Mauro, de~Mello~Neto, Mitri, de~Oliveira, de~Souza,
  Debatin, Deligny, Castro, Diogo, Dobrigkeit, D'Olivo, Dorosti, Anjos, Dova,
  Dundovic, Ebr, Engel, Erdmann, Erfani, Escobar, Espadanal, Etchegoyen,
  Falcke, Farmer, Farrar, Fauth, Fazzini, Feldbusch, Fenu, Fick, Figueira,
  Filip{\v{c}}i{\v{c}}, Freire, Fujii, Fuster, Gaïor, Garc{\'{\i}}a,
  Gat{\'{e}}, Gemmeke, Gherghel-Lascu, Ghia, Giaccari, Giammarchi, Giller,
  G{\l}as, Glaser, Golup, Berisso, Vitale, Gonz{\'{a}}lez, Gorgi, Gottowik,
  Grillo, Grubb, Guarino, Guedes, Halliday, Hampel, Hansen, Harari, Harrison,
  Harvey, Haungs, Hebbeker, Heck, Heimann, Herve, Hill, Hojvat, Holt, Homola,
  Hörandel, Horvath, Hrabovsk{\'{y}}, Huege, Hulsman, Insolia, Isar, Jandt,
  Johnsen, Josebachuili, Jurysek, Kääpä, Kampert, Keilhauer, Kemmerich,
  Kemp, Kieckhafer, Klages, Kleifges, Kleinfeller, Krause, Krohm, Kuempel,
  Mezek, Kunka, Awad, Lago, LaHurd, Lang, Lauscher, Legumina, de~Oliveira,
  Letessier-Selvon, Lhenry-Yvon, Link, Presti, Lopes, L{\'{o}}pez, Casado,
  Lorek, Luce, Lucero, Malacari, Mallamaci, Mandat, Mantsch, Mariazzi, Maris,
  Marsella, Martello, Martinez, Bravo, Meza, Mathes, Mathys, Matthews,
  Matthiae, Mayotte, Mazur, Medina, Medina-Tanco, Melo, Menshikov, Merenda,
  Michal, Micheletti, Middendorf, Miramonti, Mitrica, Mockler, Mollerach,
  Montanet, Morello, Morlino, Müller, Müller, Muller, Müller, Mussa,
  Naranjo, Nguyen, Niculescu-Oglinzanu, Niechciol, Niemietz, Niggemann, Nitz,
  Nosek, Novotny, No{\v{z}}ka, N{\'{u}}{\~{n}}ez, Oikonomou, Olinto, Palatka,
  Pallotta, Papenbreer, Parente, Parra, Paul, Pech, Pedreira, Pȩkala,
  Pe{\~{n}}a-Rodriguez, Pereira, Perlin, Perrone, Peters, Petrera, Phuntsok,
  Pierog, Pimenta, Pirronello, Platino, Plum, Poh, Porowski, Prado, Privitera,
  Prouza, Quel, Querchfeld, Quinn, Ramos-Pollan, Rautenberg, Ravignani, Ridky,
  Riehn, Risse, Ristori, Rizi, de~Carvalho, Fernandez, Rojo, Roncoroni, Roth,
  Roulet, Rovero, Ruehl, Saffi, Saftoiu, Salamida, Salazar, Saleh, Salina,
  S{\'{a}}nchez, Sanchez-Lucas, Santos, Santos, Sarazin, Sarmento,
  Sarmiento-Cano, Sato, Schauer, Scherini, Schieler, Schimp, Schmidt, Scholten,
  Schov{\'{a}}nek, Schröder, Schröder, Schulz, Schumacher, Sciutto, Segreto,
  Shadkam, Shellard, Sigl, Silli, {\v{S}}m{\'{\i}}da, Snow, Sommers, Sonntag,
  Soriano, Squartini, Stanca, Stani{\v{c}}, Stasielak, Stassi, Stolpovskiy,
  Strafella, Streich, Suarez, Suarez-Dur{\'{a}}n, Sudholz, Suomijärvi,
  Supanitsky, {\v{S}}up{\'{\i}}k, Swain, Szadkowski, Taboada, Taborda,
  Timmermans, Peixoto, Tomankova, Tom{\'{e}}, Elipe, Travnicek, Trini, Tueros,
  Ulrich, Unger, Urban, Galicia, Vali{\~{n}}o, Valore, van Aar, van Bodegom,
  van~den Berg, van Vliet, Varela, C{\'{a}}rdenas, V{\'{a}}zquez,
  Veberi{\v{c}}, Ventura, Quispe, Verzi, Vicha, Villase{\~{n}}or, Vorobiov,
  Wahlberg, Wainberg, Walz, Watson, Weber, Weindl, Wiede{\'{n}}ski, Wiencke,
  Wilczy{\'{n}}ski, Wirtz, Wittkowski, Wundheiler, Yang, Yushkov, Zas,
  Zavrtanik, Zavrtanik, Zepeda, Zimmermann, Ziolkowski, Zong, Zuccarello, Kim,
  Schulze, Bauer, Corral-Santana, de~Gregorio-Monsalvo,
  Gonz{\'{a}}lez-L{\'{o}}pez, Hartmann, Ishwara-Chandra, Mart{\'{\i}}n, Mehner,
  Misra, Micha{\l}owski, Resmi, Paragi, Agudo, An, Beswick, Casadio, Frey,
  Jonker, Kettenis, Marcote, Moldon, Szomoru, van Langevelde, Yang, Cwiek,
  Cwiok, Czyrkowski, Dabrowski, Kasprowicz, Mankiewicz, Nawrocki, Opiela,
  Piotrowski, Wrochna, Zaremba, {\.{Z}}arnecki, Haggard, Nynka, Ruan, Bland,
  Booler, Devillepoix, de~Gois, Hancock, Howie, Paxman, Sansom, Towner, Tonry,
  Coughlin, Stubbs, Denneau, Heinze, Stalder, Weiland, Eatough, Kramer, Kraus,
  Troja, Piro, Gonz{\'{a}}lez, Butler, Fox, Khandrika, Kutyrev, Lee, Ricci,
  Jr., S{\'{a}}nchez-Ram{\'{\i}}rez, Veilleux, Watson, Wieringa, Burgess, van
  Eerten, Fontes, Fryer, Korobkin, Wollaeger, Camilo, Foley, Goedhart,
  Makhathini, Oozeer, Smirnov, Fender, Woudt, , , , , , , , , , , , , , , , , ,
  , , , , , , , , , , , , , , , , , , , , , , , , , , , , , , , \&
  and}]{Abbott_2017}
Abbott, B.~P., Abbott, R., Abbott, T.~D., {et~al.} 2017, The Astrophysical
  Journal, 848, L12, \dodoi{10.3847/2041-8213/aa91c9}

\bibitem[{Abel {et~al.}(2002)Abel, Bryan, \& Norman}]{Abel93}
Abel, T., Bryan, G.~L., \& Norman, M.~L. 2002, Science, 295, 93,
  \dodoi{10.1126/science.1063991}

\bibitem[{Almusleh {et~al.}(2021)Almusleh, Taani, Özdemir, Rah, Al-Wardat,
  Zhao, \& Mardini}]{Almusleh_2021}
Almusleh, N.~A., Taani, A., Özdemir, S., {et~al.} 2021, Astronomische
  Nachrichten, 342, 625, \dodoi{https://doi.org/10.1002/asna.202113867}

\bibitem[{{Aoki} {et~al.}(2007){Aoki}, {Beers}, {Christlieb}, {Norris}, {Ryan},
  \& {Tsangarides}}]{aoki2007}
{Aoki}, W., {Beers}, T.~C., {Christlieb}, N., {et~al.} 2007, \apj, 655, 492,
  \dodoi{10.1086/509817}

\bibitem[{Asplund {et~al.}(2009)Asplund, Nicolas, Sauval, \&
  Scott}]{Asplund_2009}
Asplund, M., Nicolas, G., Sauval, A., \& Scott, P. 2009, ARA\&A, 47,
  \dodoi{10.1146/annurev.astro.46.060407.145222}

\bibitem[{{Beers} \& {Christlieb}(2005)}]{Beers_2005}
{Beers}, T.~C., \& {Christlieb}, N. 2005, \araa, 43, 531,
  \dodoi{10.1146/annurev.astro.42.053102.134057}

\bibitem[{{Beers} {et~al.}(1990){Beers}, {Flynn}, \& {Gebhardt}}]{Beers_1990}
{Beers}, T.~C., {Flynn}, K., \& {Gebhardt}, K. 1990, \aj, 100, 32,
  \dodoi{10.1086/115487}

\bibitem[{{Beers} {et~al.}(2014){Beers}, {Norris}, {Placco}, {Lee}, {Rossi},
  {Carollo}, \& {Masseron}}]{Beers_2014}
{Beers}, T.~C., {Norris}, J.~E., {Placco}, V.~M., {et~al.} 2014, \apj, 794, 58,
  \dodoi{10.1088/0004-637X/794/1/58}

\bibitem[{{Beers} {et~al.}(1992){Beers}, {Preston}, \& {Shectman}}]{Beers_1992}
{Beers}, T.~C., {Preston}, G.~W., \& {Shectman}, S.~A. 1992, \aj, 103, 1987,
  \dodoi{10.1086/116207}

\bibitem[{Beers {et~al.}(2017)Beers, Placco, Carollo, Rossi, Lee, Frebel,
  Norris, Dietz, \& Masseron}]{Beers_2017}
Beers, T.~C., Placco, V.~M., Carollo, D., {et~al.} 2017, The Astrophysical
  Journal, 835, 81, \dodoi{10.3847/1538-4357/835/1/81}

\bibitem[{Brauer {et~al.}(2019)Brauer, Ji, Frebel, Dooley, Gómez, \&
  O’Shea}]{Brauer_2019}
Brauer, K., Ji, A.~P., Frebel, A., {et~al.} 2019, The Astrophysical Journal,
  871, 247, \dodoi{10.3847/1538-4357/aafafb}

\bibitem[{Bromm {et~al.}(2002)Bromm, Coppi, \& Larson}]{Bromm_2002}
Bromm, V., Coppi, P.~S., \& Larson, R.~B. 2002, The Astrophysical Journal, 564,
  23, \dodoi{10.1086/323947}

\bibitem[{Bromm \& Loeb(2003)}]{Bromm_2003}
Bromm, V., \& Loeb, A. 2003, Nature, 425, 812, \dodoi{10.1038/nature02071}

\bibitem[{Brooke {et~al.}(2013)Brooke, Bernath, Schmidt, \&
  Bacskay}]{Brooke_2013}
Brooke, J.~S., Bernath, P.~F., Schmidt, T.~W., \& Bacskay, G.~B. 2013, Journal
  of Quantitative Spectroscopy and Radiative Transfer, 124, 11,
  \dodoi{https://doi.org/10.1016/j.jqsrt.2013.02.025}

\bibitem[{{Casey}(2014)}]{Casey2014}
{Casey}, A.~R. 2014, PhD thesis, Australian National University

\bibitem[{{Castelli} \& {Kurucz}(2003)}]{Castelli_2003}
{Castelli}, F., \& {Kurucz}, R.~L. 2003, in Modelling of Stellar Atmospheres,
  ed. N.~{Piskunov}, W.~{Weiss}, \& D.~{Gray}, Vol. 210

\bibitem[{Chiaki {et~al.}(2017)Chiaki, Tominaga, \& Nozawa}]{Chiaki_2017}
Chiaki, G., Tominaga, N., \& Nozawa, T. 2017, Monthly Notices of the Royal
  Astronomical Society: Letters, 472, L115, \dodoi{10.1093/mnrasl/slx163}

\bibitem[{Chiaki {et~al.}(2020)Chiaki, Wise, Marassi, Schneider, Limongi, \&
  Chieffi}]{Chiaki_2020}
Chiaki, G., Wise, J.~H., Marassi, S., {et~al.} 2020, Monthly Notices of the
  Royal Astronomical Society, 497, 3149, \dodoi{10.1093/mnras/staa2144}

\bibitem[{{Choplin} {et~al.}(2016){Choplin}, {Maeder Andr\'e}, {Meynet
  Georges}, \& {Chiappini Cristina}}]{Choplin_2016}
{Choplin}, {Maeder Andr\'e}, {Meynet Georges}, \& {Chiappini Cristina}. 2016,
  A\&A, 593, A36, \dodoi{10.1051/0004-6361/201628083}

\bibitem[{Cooke {et~al.}(2011{\natexlab{a}})Cooke, Pettini, Steidel, Rudie, \&
  Jorgenson}]{Cooke:2011au}
Cooke, R., Pettini, M., Steidel, C.~C., Rudie, G.~C., \& Jorgenson, R.~A.
  2011{\natexlab{a}}, Mon. Not. Roy. Astron. Soc., 412, 1047,
  \dodoi{10.1111/j.1365-2966.2010.17966.x}

\bibitem[{Cooke {et~al.}(2011{\natexlab{b}})Cooke, Pettini, Steidel, Rudie, \&
  Nissen}]{10.1111/j.1365-2966.2011.19365.x}
Cooke, R., Pettini, M., Steidel, C.~C., Rudie, G.~C., \& Nissen, P.~E.
  2011{\natexlab{b}}, Monthly Notices of the Royal Astronomical Society, 417,
  1534, \dodoi{10.1111/j.1365-2966.2011.19365.x}

\bibitem[{Cowan {et~al.}(2021)Cowan, Sneden, Lawler, Aprahamian, Wiescher,
  Langanke, Mart\'{\i}nez-Pinedo, \& Thielemann}]{Cowan_2021}
Cowan, J.~J., Sneden, C., Lawler, J.~E., {et~al.} 2021, Rev. Mod. Phys., 93,
  015002, \dodoi{10.1103/RevModPhys.93.015002}

\bibitem[{Demarque {et~al.}(2004)Demarque, Woo, Kim, \& Yi}]{Demarque_2004}
Demarque, P., Woo, J.-H., Kim, Y.-C., \& Yi, S.~K. 2004, The Astrophysical
  Journal Supplement Series, 155, 667, \dodoi{10.1086/424966}

\bibitem[{Drout {et~al.}(2017)Drout, Piro, Shappee, Kilpatrick, Simon,
  Contreras, Coulter, Foley, Siebert, Morrell, Boutsia, Mille, Holoien, Kasen,
  Kollmeier, Madore, Monson, Murguia-Berthier, Pan, Prochaska, Ramirez-Ruiz,
  Rest, Adams, Alatalo, Bañados, Baughman, Beers, Bernstein, Bitsakis,
  Campillay, Hansen, Higgs, Ji, Maravelias, Marshall, Bidin, Prieto, Rasmussen,
  Rojas-Bravo, Strom, Ulloa, Vargas-González, Wan, \& Whitten}]{Drout_2017}
Drout, M.~R., Piro, A.~L., Shappee, B.~J., {et~al.} 2017, Science, 358, 1570,
  \dodoi{10.1126/science.aaq0049}

\bibitem[{Dutta {et~al.}(2020)Dutta, Sur, Stacy, \& Bagla}]{Dutta_2020}
Dutta, J., Sur, S., Stacy, A., \& Bagla, J.~S. 2020, The Astrophysical Journal,
  901, 16, \dodoi{10.3847/1538-4357/abadf8}

\bibitem[{Frebel(2018)}]{Frebel_2018}
Frebel, A. 2018, Annual Review of Nuclear and Particle Science, 68, 237,
  \dodoi{10.1146/annurev-nucl-101917-021141}

\bibitem[{Frebel \& Bromm(2012)}]{Frebel_2012}
Frebel, A., \& Bromm, V. 2012, The Astrophysical Journal, 759, 115,
  \dodoi{10.1088/0004-637x/759/2/115}

\bibitem[{{Frebel} {et~al.}(2013){Frebel}, {Casey}, {Jacobson}, \&
  {Yu}}]{Frebel2013}
{Frebel}, A., {Casey}, A.~R., {Jacobson}, H.~R., \& {Yu}, Q. 2013, \apj, 769,
  57, \dodoi{10.1088/0004-637X/769/1/57}

\bibitem[{{Frebel} {et~al.}(2007){Frebel}, {Johnson}, \& {Bromm}}]{Frebel_2007}
{Frebel}, A., {Johnson}, J.~L., \& {Bromm}, V. 2007, \mnras, 380, L40,
  \dodoi{10.1111/j.1745-3933.2007.00344.x}

\bibitem[{{Frebel} \& {Norris}(2015)}]{Frebel_2015}
{Frebel}, A., \& {Norris}, J.~E. 2015, \araa, 53, 631,
  \dodoi{10.1146/annurev-astro-082214-122423}

\bibitem[{{Gallagher} {et~al.}(2010){Gallagher}, {Ryan}, {Garc\'{\i}a}~P\'erez,
  \& {Aoki}}]{Gallagher_2010}
{Gallagher}, A.~J., {Ryan}, S.~G., {Garc\'{\i}a}~P\'erez, A.~E., \& {Aoki}, W.
  2010, A\&A, 523, A24, \dodoi{10.1051/0004-6361/201014970}

\bibitem[{{Goswami} {et~al.}(2021){Goswami}, {Rathour, Rajeev Singh}, \&
  {Goswami, Aruna}}]{Goswami_2021}
{Goswami}, P.~P., {Rathour, Rajeev Singh}, \& {Goswami, Aruna}. 2021, A\&A,
  649, A49, \dodoi{10.1051/0004-6361/202038258}

\bibitem[{Gudin {et~al.}(2021)Gudin, Shank, Beers, Yuan, Limberg, Roederer,
  Placco, Holmbeck, Dietz, Rasmussen, Hansen, Sakari, Ezzeddine, \&
  Frebel}]{Gudin_2021}
Gudin, D., Shank, D., Beers, T.~C., {et~al.} 2021, The Astrophysical Journal,
  908, 79, \dodoi{10.3847/1538-4357/abd7ed}

\bibitem[{Hampel {et~al.}(2016)Hampel, Stancliffe, Lugaro, \&
  Meyer}]{Hampel_2016}
Hampel, M., Stancliffe, R.~J., Lugaro, M., \& Meyer, B.~S. 2016, The
  Astrophysical Journal, 831, 171, \dodoi{10.3847/0004-637x/831/2/171}

\bibitem[{Hansen {et~al.}(2017)}]{Hansen_2017}
Hansen, T.~T., {et~al.} 2017, Astrophys. J., 838, 44,
  \dodoi{10.3847/1538-4357/aa634a}

\bibitem[{Hansen {et~al.}(2018)Hansen, Holmbeck, Beers, Placco, Roederer,
  Frebel, Sakari, Simon, \& Thompson}]{Hansen2018}
Hansen, T.~T., Holmbeck, E.~M., Beers, T.~C., {et~al.} 2018, \apj, 858, 92,
  \dodoi{10.3847/1538-4357/aabacc}

\bibitem[{Holmbeck {et~al.}(2020)Holmbeck, Hansen, Beers, Placco, Whitten,
  Rasmussen, Roederer, Ezzeddine, Sakari, Frebel, Drout, Simon, Thompson,
  Bland-Hawthorn, Gibson, Grebel, Kordopatis, Kunder, Mel{\'{e}}ndez, Navarro,
  Reid, Seabroke, Steinmetz, Watson, \& Wyse}]{Holmbeck_2020}
Holmbeck, E.~M., Hansen, T.~T., Beers, T.~C., {et~al.} 2020, The Astrophysical
  Journal Supplement Series, 249, 30, \dodoi{10.3847/1538-4365/ab9c19}

\bibitem[{Huang {et~al.}(2021)Huang, Beers, Wolf, Lee, Onken, Yuan, Shank,
  Zhang, Wang, Shi, \& Fan}]{Huang_2021}
Huang, Y., Beers, T., Wolf, C., {et~al.} 2021, Beyond spectroscopy. I.
  Metallicities, distances, and age estimates for over twenty million stars
  from SMSS DR2 and Gaia EDR3

\bibitem[{Ishigaki {et~al.}(2014)Ishigaki, Tominaga, Kobayashi, \&
  Nomoto}]{Ishigaki_2014}
Ishigaki, M.~N., Tominaga, N., Kobayashi, C., \& Nomoto, K. 2014, The
  Astrophysical Journal, 792, L32, \dodoi{10.1088/2041-8205/792/2/l32}

\bibitem[{Jacobson {et~al.}(2015)Jacobson, Keller, Frebel, Casey, Asplund,
  Bessell, Costa, Lind, Marino, Norris, Pe{\~{n}}a, Schmidt, Tisserand, Walsh,
  Yong, \& Yu}]{Jacobson2015f}
Jacobson, H.~R., Keller, S., Frebel, A., {et~al.} 2015, \apj, 807, 171,
  \dodoi{10.1088/0004-637X/807/2/171}

\bibitem[{{Ji} {et~al.}(2016){Ji}, {Frebel}, {Chiti}, \& {Simon}}]{Ji_2016}
{Ji}, A.~P., {Frebel}, A., {Chiti}, A., \& {Simon}, J.~D. 2016, \nat, 531, 610,
  \dodoi{10.1038/nature17425}

\bibitem[{{Ji} {et~al.}(2019){Ji}, {Simon}, {Frebel}, {Venn}, \&
  {Hansen}}]{Ji_2019}
{Ji}, A.~P., {Simon}, J.~D., {Frebel}, A., {Venn}, K.~A., \& {Hansen}, T.~T.
  2019, \apj, 870, 83, \dodoi{10.3847/1538-4357/aaf3bb}

\bibitem[{Kilpatrick {et~al.}(2017)Kilpatrick, Foley, Kasen, Murguia-Berthier,
  Ramirez-Ruiz, Coulter, Drout, Piro, Shappee, Boutsia, Contreras, Mille,
  Madore, Morrell, Pan, Prochaska, Rest, Rojas-Bravo, Siebert, Simon, \&
  Ulloa}]{Kilpatrick_2017}
Kilpatrick, C.~D., Foley, R.~J., Kasen, D., {et~al.} 2017, Science, 358, 1583,
  \dodoi{10.1126/science.aaq0073}

\bibitem[{Kobayashi {et~al.}(2020)Kobayashi, Karakas, \&
  Lugaro}]{Kobayashi_2020}
Kobayashi, C., Karakas, A.~I., \& Lugaro, M. 2020, The Astrophysical Journal,
  900, 179, \dodoi{10.3847/1538-4357/abae65}

\bibitem[{{Kobayashi} \& {Nomoto}(2009)}]{Kobayashi_2009}
{Kobayashi}, C., \& {Nomoto}, K. 2009, \apj, 707, 1466,
  \dodoi{10.1088/0004-637X/707/2/1466}

\bibitem[{Kunder {et~al.}(2017)Kunder, Kordopatis, Steinmetz, Zwitter,
  McMillan, Casagrande, Enke, Wojno, Valentini, Chiappini, Matijevi{\v{c}},
  Siviero, de~Laverny, Recio-Blanco, Bijaoui, Wyse, Binney, Grebel, Helmi,
  Jofre, Antoja, Gilmore, Siebert, Famaey, Bienaym{\'{e}}, Gibson, Freeman,
  Navarro, Munari, Seabroke, Anguiano, {\v{Z}}erjal, Minchev, Reid,
  Bland-Hawthorn, Kos, Sharma, Watson, Parker, Scholz, Burton, Cass, Hartley,
  Fiegert, Stupar, Ritter, Hawkins, Gerhard, Chaplin, Davies, Elsworth, Lund,
  Miglio, \& Mosser}]{Rave5}
Kunder, A., Kordopatis, G., Steinmetz, M., {et~al.} 2017, The Astronomical
  Journal, 153, 75, \dodoi{10.3847/1538-3881/153/2/75}

\bibitem[{Lawler {et~al.}(2001)Lawler, Bonvallet, \& Sneden}]{Lawler_2001}
Lawler, J.~E., Bonvallet, G., \& Sneden, C. 2001, The Astrophysical Journal,
  556, 452, \dodoi{10.1086/321549}

\bibitem[{Lee {et~al.}(2013)Lee, Beers, Masseron, Plez, Rockosi, Sobeck, Yanny,
  Lucatello, Sivarani, Placco, \& Carollo}]{Lee_2013}
Lee, Y.~S., Beers, T.~C., Masseron, T., {et~al.} 2013, The Astronomical
  Journal, 146, 132, \dodoi{10.1088/0004-6256/146/5/132}

\bibitem[{Lucatello {et~al.}(2006)Lucatello, Beers, Christlieb, Barklem, Rossi,
  Marsteller, Sivarani, \& Lee}]{Lucatello_2006}
Lucatello, S., Beers, T.~C., Christlieb, N., {et~al.} 2006, The Astrophysical
  Journal, 652, L37, \dodoi{10.1086/509780}

\bibitem[{Marshall {et~al.}(2019)Marshall, Hansen, Simon, Li, Bernstein, Kuehn,
  Pace, DePoy, Palmese, Pieres, Strigari, Drlica-Wagner, Bechtol, Lidman,
  Nagasawa, Bertin, Brooks, Buckley-Geer, Burke, Rosell, Kind, Carretero,
  Cunha, D'Andrea, da~Costa, Vicente, Desai, Doel, Eifler, Flaugher, Fosalba,
  Frieman, Garc{\'{\i}}a-Bellido, Gaztanaga, Gerdes, Gruendl, Gschwend,
  Gutierrez, Hartley, Hollowood, Honscheid, Hoyle, James, Kuropatkin, Maia,
  Menanteau, Miller, Miquel, Plazas, Sanchez, Santiago, Scarpine, Schubnell,
  Serrano, Sevilla-Noarbe, Smith, Soares-Santos, Suchyta, Swanson, Tarle, \&
  and}]{Marshall_2019}
Marshall, J.~L., Hansen, T., Simon, J.~D., {et~al.} 2019, The Astrophysical
  Journal, 882, 177, \dodoi{10.3847/1538-4357/ab3653}

\bibitem[{Masseron {et~al.}(2014)Masseron, Plez, Van~Eck, Colin, Daoutidis,
  Godefroid, Coheur, Bernath, Jorissen, \& Christlieb}]{Masseron_2014}
Masseron, T., Plez, B., Van~Eck, S., {et~al.} 2014, A\&A, 571, A47,
  \dodoi{10.1051/0004-6361/201423956}

\bibitem[{Meynet {et~al.}(2006)Meynet, Ekstr\"om, \& Maeder}]{Meynet_2006}
Meynet, G., Ekstr\"om, S., \& Maeder, A. 2006, A\&A, 447, 623,
  \dodoi{10.1051/0004-6361:20053070}

\bibitem[{Meynet {et~al.}(2010)Meynet, Hirschi, Ekstr\"om, Maeder, Georgy,
  Eggenberger, \& Chiappini}]{Meynet_2010}
Meynet, G., Hirschi, R., Ekstr\"om, S., {et~al.} 2010, A\&A, 521, A30,
  \dodoi{10.1051/0004-6361/200913377}

\bibitem[{Nomoto {et~al.}(2006)Nomoto, Tominaga, Umeda, Kobayashi, \&
  Maeda}]{Nomoto_2006}
Nomoto, K., Tominaga, N., Umeda, H., Kobayashi, C., \& Maeda, K. 2006, Nuclear
  Physics A, 777, 424, \dodoi{https://doi.org/10.1016/j.nuclphysa.2006.05.008}

\bibitem[{Peebles(1966)}]{PhysRevLett.16.410}
Peebles, P. J.~E. 1966, Phys. Rev. Lett., 16, 410,
  \dodoi{10.1103/PhysRevLett.16.410}

\bibitem[{Placco {et~al.}(2014)Placco, Frebel, Beers, \&
  Stancliffe}]{Placco_2014}
Placco, V.~M., Frebel, A., Beers, T.~C., \& Stancliffe, R.~J. 2014, The
  Astrophysical Journal, 797, 21, \dodoi{10.1088/0004-637x/797/1/21}

\bibitem[{{Placco} {et~al.}(2021){Placco}, {Sneden}, {Roederer}, {Lawler}, {Den
  Hartog}, {Hejazi}, {Maas}, \& {Bernath}}]{Placco_2021}
{Placco}, V.~M., {Sneden}, C., {Roederer}, I.~U., {et~al.} 2021, Research Notes
  of the American Astronomical Society, 5, 92, \dodoi{10.3847/2515-5172/abf651}

\bibitem[{Placco {et~al.}(2018)Placco, Beers, Santucci, Chanam{\'{e}},
  Sep{\'{u}}lveda, Coronado, Points, Kaleida, Rossi, Kordopatis, Lee,
  Matijevi{\v{c}}, Frebel, Hansen, Holmbeck, Rasmussen, Roederer, Sakari, \&
  Whitten}]{Placco_2018}
Placco, V.~M., Beers, T.~C., Santucci, R.~M., {et~al.} 2018, The Astronomical
  Journal, 155, 256, \dodoi{10.3847/1538-3881/aac20c}

\bibitem[{Purandardas \& Goswami(2021)}]{Purandardas_2021}
Purandardas, M., \& Goswami, A. 2021, Observational evidence points at AGB
  stars as possible progenitors of CEMP-s \& r/s stars.
\newblock \doarXiv{2108.06075}

\bibitem[{Ram {et~al.}(2014)Ram, Brooke, Bernath, Sneden, \&
  Lucatello}]{Ram_2014}
Ram, R.~S., Brooke, J. S.~A., Bernath, P.~F., Sneden, C., \& Lucatello, S.
  2014, The Astrophysical Journal Supplement Series, 211, 5,
  \dodoi{10.1088/0067-0049/211/1/5}

\bibitem[{Rasmussen {et~al.}(2020)Rasmussen, Zepeda, Beers, Placco, Depagne,
  Frebel, Dietz, \& Hartwig}]{Rasmussen_2020}
Rasmussen, K.~C., Zepeda, J., Beers, T.~C., {et~al.} 2020, The Astrophysical
  Journal, 905, 20, \dodoi{10.3847/1538-4357/abc005}

\bibitem[{Reggiani {et~al.}(2021)Reggiani, Schlaufman, Casey, Simon, \&
  Ji}]{Reggiani_2021}
Reggiani, H., Schlaufman, K.~C., Casey, A.~R., Simon, J.~D., \& Ji, A.~P. 2021,
  The Most Metal-poor Stars in the Magellanic Clouds are $r$-process Enhanced.
\newblock \doarXiv{2108.10880}

\bibitem[{{Roederer} {et~al.}(2018){Roederer}, {Hattori}, \&
  {Valluri}}]{Roederer_2018}
{Roederer}, I.~U., {Hattori}, K., \& {Valluri}, M. 2018, \aj, 156, 179,
  \dodoi{10.3847/1538-3881/aadd9c}

\bibitem[{{Roederer} {et~al.}(2016){Roederer}, {Mateo}, {Bailey}, {Song},
  {Bell}, {Crane}, {Loebman}, {Nidever}, {Olszewski}, {Shectman}, {Thompson},
  {Valluri}, \& {Walker}}]{Roederer_2016}
{Roederer}, I.~U., {Mateo}, M., {Bailey}, John~I., I., {et~al.} 2016, \aj, 151,
  82, \dodoi{10.3847/0004-6256/151/3/82}

\bibitem[{{Rossi} {et~al.}(1999){Rossi}, {Beers}, \& {Sneden}}]{Rossi_1999}
{Rossi}, S., {Beers}, T.~C., \& {Sneden}, C. 1999, in Astronomical Society of
  the Pacific Conference Series, Vol. 165, The Third Stromlo Symposium: The
  Galactic Halo, ed. B.~K. {Gibson}, R.~S. {Axelrod}, \& M.~E. {Putman}, 264

\bibitem[{Rossi {et~al.}(2005)Rossi, Beers, Sneden, Sevastyanenko, Rhee, \&
  Marsteller}]{Rossi_2005}
Rossi, S., Beers, T.~C., Sneden, C., {et~al.} 2005, The Astronomical Journal,
  130, 2804, \dodoi{10.1086/497164}

\bibitem[{{Sabano} \& {Yoshii}(1977)}]{Sabano_1977}
{Sabano}, Y., \& {Yoshii}, Y. 1977, \pasj, 29, 207

\bibitem[{{Sakari} {et~al.}(2018){Sakari}, {Placco}, {Farrell}, {Roederer},
  {Wallerstein}, {Beers}, {Ezzeddine}, {Frebel}, {Hansen}, {Holmbeck},
  {Sneden}, {Cowan}, {Venn}, {Davis}, {Matijevi{\v{c}}}, {Wyse},
  {Bland-Hawthorn}, {Chiappini}, {Freeman}, {Gibson}, {Grebel}, {Helmi},
  {Kordopatis}, {Kunder}, {Navarro}, {Reid}, {Seabroke}, {Steinmetz}, \&
  {Watson}}]{Sakari2018b}
{Sakari}, C.~M., {Placco}, V.~M., {Farrell}, E.~M., {et~al.} 2018, \apj, 868,
  110, \dodoi{10.3847/1538-4357/aae9df}

\bibitem[{Shappee {et~al.}(2017)Shappee, Simon, Drout, Piro, Morrell, Prieto,
  Kasen, Holoien, Kollmeier, Kelson, Coulter, Foley, Kilpatrick, Siebert,
  Madore, Murguia-Berthier, Pan, Prochaska, Ramirez-Ruiz, Rest, Adams, Alatalo,
  Bañados, Baughman, Bernstein, Bitsakis, Boutsia, Bravo, Mille, Higgs, Ji,
  Maravelias, Marshall, Placco, Prieto, \& Wan}]{Shappee_2017}
Shappee, B.~J., Simon, J.~D., Drout, M.~R., {et~al.} 2017, Science, 358, 1574,
  \dodoi{10.1126/science.aaq0186}

\bibitem[{{Silk}(1977)}]{Silk_1977}
{Silk}, J. 1977, \apj, 211, 638, \dodoi{10.1086/154972}

\bibitem[{{Sneden}(1973)}]{Sneden_1973}
{Sneden}, C. 1973, \apj, 184, 839, \dodoi{10.1086/152374}

\bibitem[{Sneden {et~al.}(2014)Sneden, Lucatello, Ram, Brooke, \&
  Bernath}]{Sneden_2014}
Sneden, C., Lucatello, S., Ram, R.~S., Brooke, J. S.~A., \& Bernath, P. 2014,
  The Astrophysical Journal Supplement Series, 214, 26,
  \dodoi{10.1088/0067-0049/214/2/26}

\bibitem[{Sobeck {et~al.}(2011)Sobeck, Kraft, Sneden, Preston, Cowan, Smith,
  Thompson, Shectman, \& Burley}]{Sobeck_2011}
Sobeck, J.~S., Kraft, R.~P., Sneden, C., {et~al.} 2011, 141, 175,
  \dodoi{10.1088/0004-6256/141/6/175}

\bibitem[{{Steinmetz} {et~al.}(2006){Steinmetz}, {Zwitter}, {Siebert},
  {Watson}, {Freeman}, {Munari}, {Campbell}, {Williams}, {Seabroke}, {Wyse},
  {Parker}, {Bienaym{\'e}}, {Roeser}, {Gibson}, {Gilmore}, {Grebel}, {Helmi},
  {Navarro}, {Burton}, {Cass}, {Dawe}, {Fiegert}, {Hartley}, {Russell},
  {Saunders}, {Enke}, {Bailin}, {Binney}, {Bland-Hawthorn}, {Boeche}, {Dehnen},
  {Eisenstein}, {Evans}, {Fiorucci}, {Fulbright}, {Gerhard}, {Jauregi}, {Kelz},
  {Mijovi{\'c}}, {Minchev}, {Parmentier}, {Pe{\~n}arrubia}, {Quillen}, {Read},
  {Ruchti}, {Scholz}, {Siviero}, {Smith}, {Sordo}, {Veltz}, {Vidrih}, {von
  Berlepsch}, {Boyle}, \& {Schilbach}}]{Rave1}
{Steinmetz}, M., {Zwitter}, T., {Siebert}, A., {et~al.} 2006, \aj, 132, 1645,
  \dodoi{10.1086/506564}

\bibitem[{{Suda} {et~al.}(2011){Suda}, {Yamada}, {Katsuta}, {Komiya},
  {Ishizuka}, {Aoki}, \& {Fujimoto}}]{Saga}
{Suda}, T., {Yamada}, S., {Katsuta}, Y., {et~al.} 2011, \mnras, 412, 843,
  \dodoi{10.1111/j.1365-2966.2011.17943.x}

\bibitem[{Susa(2019)}]{Susa_2019}
Susa, H. 2019, The Astrophysical Journal, 877, 99,
  \dodoi{10.3847/1538-4357/ab1b6f}

\bibitem[{Susa {et~al.}(2014)Susa, Hasegawa, \& Tominaga}]{Susa_2014}
Susa, H., Hasegawa, K., \& Tominaga, N. 2014, The Astrophysical Journal, 792,
  32, \dodoi{10.1088/0004-637x/792/1/32}

\bibitem[{{Tominaga} {et~al.}(2014){Tominaga}, {Iwamoto}, \&
  {Nomoto}}]{Tominaga_2014}
{Tominaga}, N., {Iwamoto}, N., \& {Nomoto}, K. 2014, \apj, 785, 98,
  \dodoi{10.1088/0004-637X/785/2/98}

\bibitem[{{Wagoner} {et~al.}(1967){Wagoner}, {Fowler}, \&
  {Hoyle}}]{Wagoner_1967}
{Wagoner}, R.~V., {Fowler}, W.~A., \& {Hoyle}, F. 1967, \apj, 148, 3,
  \dodoi{10.1086/149126}

\bibitem[{Yoon {et~al.}(2016)Yoon, Beers, Placco, Rasmussen, Carollo, He,
  Hansen, Roederer, \& Zeanah}]{Yoon_2016}
Yoon, J., Beers, T.~C., Placco, V.~M., {et~al.} 2016, The Astrophysical
  Journal, 833, 20, \dodoi{10.3847/0004-637x/833/1/20}

\bibitem[{Yoon {et~al.}(2018)Yoon, Beers, Dietz, Lee, Placco, Costa, Keller,
  Owen, \& Sharma}]{Yoon_2018}
Yoon, J., Beers, T.~C., Dietz, S., {et~al.} 2018, The Astrophysical Journal,
  861, 146, \dodoi{10.3847/1538-4357/aaccea}

\bibitem[{Zou {et~al.}(2020)Zou, Petitjean, Noterdaeme, Ledoux, Srianand,
  Jiang, \& Krogager}]{Zou_2020}
Zou, S., Petitjean, P., Noterdaeme, P., {et~al.} 2020, The Astrophysical
  Journal, 901, 105, \dodoi{10.3847/1538-4357/abb092}

\end{thebibliography}

\end{document}